\documentclass[preprint,12pt,3p,authoryear,letterpaper,top=2cm,bottom=2cm,left=3cm,right=3cm,marginparwidth=1.75cm]{elsarticle}

\usepackage[english]{babel}

\usepackage{geometry}
\usepackage{xcolor}
\usepackage{capt-of}
\usepackage{caption}
\usepackage{aas_macros}
\usepackage{amsmath}
\usepackage{amssymb} 
\usepackage{float}
\usepackage{graphicx}
\usepackage[colorlinks=true, allcolors=blue]{hyperref}

\usepackage{natbib}
\usepackage{booktabs}
\usepackage{array}
\newcolumntype{R}[1]{>{\raggedleft\arraybackslash}p{#1}}
\usepackage{siunitx}
\usepackage{tablefootnote}
\usepackage{footmisc}

\usepackage{longtable}
\usepackage{booktabs} 

\setcitestyle{authoryear,open={(},close={)}}

\newcommand{\tss}[1]{{\textsuperscript{#1}}}
\newcommand{\fig}[2]{{fig.\,\ref{#1}{#2}}}
\newcommand{\tab}[1]{{table\,\ref{#1}}}
\newcommand{\eqn}[1]{{eq.\,\ref{#1}}}

\newcommand{\cf}{{cf.}}
\newcommand{\eg}{{e.g.}}
\newcommand{\ie}{{i.e.}}

\newcommand{\dif}{\mathrm{d}}
\newcommand{\aei}{(a,e,i)}

\newcommand{\deltav}{\Delta v}

\newcommand{\arcdeg}{{^{\circ}}}

\newcommand{\au}{\,\mathrm{au}}
\newcommand{\km}{\,\mathrm{km}}

\newcommand{\meter}{\,\mathrm{m}}

\newcommand{\yr}{\,\mathrm{yr}}
\newcommand{\Myr}{\,\mathrm{Myr}}

\newcommand{\Days}{\,\mathrm{days}}

\newcommand{\second}{\,\mathrm{s}}
\newcommand{\mps}{\,\meter\,\second^{-1}}

\newcommand{\namedasteroid}[2]{{({#1})\,{#2}}}

\newcommand{\designation}[2]{{{#1}\,{#2}}}
\newcommand{\designationsub}[3]{{{#1}\,{#2}{$_{#3}$}}}

\newcommand{\CD}{{2020\,CD$_3$}}
\newcommand{\PT}{{2024\,PT$_5$}}

\newcommand{\Kamo}{{\namedasteroid{469219}{Kamo‘oalewa}}}

\newcommand{\PS}{\protect \hbox {Pan-STARRS}}

\newcommand{\nKnownCoorbitals}{97}
\newcommand{\fAvgInclination}{7.5}

\newcommand{\fMedianAbsMag}{26.4}
\newcommand{\nKnownCoorbitalsTenM}{76}
\newcommand{\fAvgEccentricity}{0.19}
\newcommand{\fMedianDiameterMeters}{19}
\newcommand{\fLargestDiameterMeters}{3000}
\newcommand{\fAvgSemiMajorAxisAU}{1.001}

\newcommand{\nNEOMODgenerated}{67,775}
\newcommand{\nNEOMODcoorb}{1589}
\newcommand{\nNEOMODcoorbAPPROX}{1600}
\newcommand{\nKnownCoorbitalsNOW}{67}
\newcommand{\nKnownCoorbitalsNOWTenM}{57}
\newcommand{\nKnownCoorbitalsNOWHundredM}{12}
\newcommand{\nNEOMODcoorbHundredM}{6}
\newcommand{\fNEOMODavgE}{0.28}
\newcommand{\fNEOMODavgI}{15.9}
\newcommand{\synLunarMaxe}{0.55}
\newcommand{\synLunarMaxi}{29}
\newcommand{\nCumulativeCoorbitalsTenMeterANYAPPROX}{70}
\newcommand{\nCumulativeCoorbitalsTenMeterANY}{71}
\newcommand{\nCumulativeCoorbitalsTenMeterQS}{37}
\newcommand{\nCumulativeCoorbitalsTenMeterHS}{64}
\newcommand{\nCumulativeCoorbitalsTenMeterTP}{18}
\newcommand{\nCumulativeCoorbitalsTenMeterCP}{17}

\newcommand{\nKnownCoorbNOWANY}{67}
\newcommand{\nKnownCoorbNOWANYTenMeter}{57}
\newcommand{\nKnownCoorbNOWANYHTwoSix}{37}
\newcommand{\nKnownCoorbNOWQS}{13}
\newcommand{\nKnownCoorbNOWQSTenMeter}{12}
\newcommand{\nKnownCoorbNOWQSHTwoSix}{6}
\newcommand{\nKnownCoorbNOWHS}{17}
\newcommand{\nKnownCoorbNOWHSTenMeter}{15}
\newcommand{\nKnownCoorbNOWHSHTwoSix}{9}
\newcommand{\nKnownCoorbNOWTP}{5}
\newcommand{\nKnownCoorbNOWTPTenMeter}{5}
\newcommand{\nKnownCoorbNOWTPHTwoSix}{5}
\newcommand{\nKnownCoorbNOWCP}{32}
\newcommand{\nKnownCoorbNOWCPTenMeter}{25}
\newcommand{\nKnownCoorbNOWCPHTwoSix}{17}
\newcommand{\LunarCoorbitalPercentMAX}{58}
\newcommand{\LunarCoorbitalPercentKamo}{21}

\newcommand{\nCoorbitalsLikeKamo}{44}
\newcommand{\nQuasiSatellitesLikeKamo}{1}

\begin{document}
\begin{frontmatter}

\title{The steady-state population of Earth's co-orbitals\\of lunar provenance}

\author[label1]{Elisa Maria Alessi}
%\ead{elisamaria.alessi@cnr.it}
\author[label2]{Robert Jedicke}
%\ead{jedicke@hawaii.edu}
\address[label1]{Istituto di Matematica Applicata e Tecnologie Informatiche, Consiglio Nazionale delle Ricerche (IMATI-CNR), Via Alfonso Corti 12, 20133 Milano, Italy. elisamaria.alessi@cnr.it}
\address[label2]{Institute for Astronomy, University of Hawai`i, 2680 Woodlawn Drive, Honolulu, HI 96822, USA. jedicke@hawaii.edu}

\begin{abstract}
The population of natural objects in a 1:1 mean motion resonance with Earth are known as Earth's co-orbitals.  Main belt objects can dynamically evolve into Earth co-orbitals but taxonomic studies of some of them have suggested that they are more likely to be lunar material.  While it has long been known that lunar ejecta can achieve Earth co-orbital status, in this work we calculate their expected steady-state size-frequency distribution from the impact rate of asteroids and comets on the Moon's surface, the ejecta's size-frequency and speed distribution, and dynamical integration of the particles for millions of years, among other factors.  We also classify known and synthetic co-orbitals by their regime (quasi-satellite, horseshoe, tadpole, or compound) and compute the probability of transitions between them.  Our nominal solution predicts that there are $\gtrsim\nCumulativeCoorbitalsTenMeterANYAPPROX$ Earth co-orbitals in the steady-state population larger than $10\meter$ in diameter with a lunar provenance but there are orders-of-magnitude systematic uncertainty on the value.  We used NEOMOD3 to calculate that about $\nNEOMODcoorbAPPROX$ are expected in the co-orbital population with a main belt provenance and they have higher eccentricity and inclination than those from the Moon.  New taxonomic classifications for more Earth co-orbitals will reduce the uncertainties on \eg\ crater scaling relations that will, in turn, reduce the uncertainties in the calculation of the steady-state population of Earth's co-orbitals with a lunar origin.  The mineralogy and abundance of Earth's co-orbitals is also of interest to commercial asteroid mining ventures because they are the lowest $\deltav$ targets in the asteroid population.
\end{abstract}

\begin{keyword}
% from https://legacyfileshare.elsevier.com/promis_misc/yicar_key_words_may2013.pdf
%%Asteroids \sep Asteroids, dynamics \sep Cratering \sep %%Moon \sep Near-Earth objects \sep Satellites, general
\end{keyword}

\end{frontmatter}

%##################################################

\section{Introduction}

When two objects orbit a primary body in the same amount of time, \ie\ have nearly identical semi-major axes, they are in a 1:1 mean motion resonance (MMR), or {\it co-orbital motion}, like Jupiter's Trojans and some moons, \eg\ Janus and Epimetheus \citep{MoraisMorbidelli2002}.  The Earth is accompanied by several known co-orbitals, the first one identified as such being \namedasteroid{3753}{Cruithne} \citep{Wiegert1997Cruithne}, and including amongst its members \Kamo, the target of the China National Space Administration's Tianwen-2 spacecraft mission \citep[\eg][]{Zhang2025}.  Dynamical integration of main belt objects into the inner solar system suggest that co-orbitals can be supplied by the main belt \citep[\eg][]{Granvik2018-NEOmodel, Nesvorny2024-NEOMOD3} but the colors and spectra of several co-orbitals are more consistent with lunar material \citep[\eg][]{Bolin2020-CD3, Sharkey2021-2016HO3,Kareta2025-PT5}.  Thus, while it has long been known that lunar ejecta can become co-orbital, in this work we calculate the expected steady-state size-frequency distribution of Earth's co-orbitals of lunar provenance.

Co-orbitals can be classified into three regimes in the prograde quasi-planar scenario:  tadpole, horseshoe, and quasi-satellite. In the framework of the circular restricted three-body problem (CR3B), tadpole orbits are Lyapunov-stable, stemming from the $L_4$ and $L_5$ equilibrium points, and quasi-periodic in their neighborhood \cite[\eg][]{Sze67, GoJoSiMas2001}, horseshoe orbits correspond to the hyperbolic invariant manifold of the $L_3$ point and to families of stable and unstable periodic and quasi-periodic orbits in its neighborhood  \citep{BaMi,2006BaOl}, and quasi-satellite orbits appear to orbit the planet, the secondary body, in a retrograde manner.

The tadpole regime has been studied extensively, especially in the framework of the Trojan asteroids of the Sun-Jupiter system \citep[\eg][]{MikkolaEtAl1994,RobGabJor,Hui, Li2023, Greenstreet}.  Every planet in our solar system except Mercury \citep[\eg][]{MarsTroj1, MarsTroj2, VenusTroj, SatTroj, UranusTroj, NeptTroj} has at least one associated Trojan asteroid and Earth has two, 2010 TK$_7$ and 2020 XL$_5$ \citep{Hui,2022Santana}. 

The quasi-satellite family of periodic and quasi-periodic orbits is of interest because it arises in correspondence with the `f' family of periodic orbits  in the Str\"omgen classification \citep{Str1933}, what Poincar\'e called `a family of the first kind' \citep{Sze67}. This means that we can generate the first orbit of the family starting from a circular periodic orbit in the Keplerian approximation and perturb it by introducing a secondary massive body into the system. The additional source of gravity does not change the stability of the family of periodic orbits \citep[\cf][and references therein]{MikInn, 2017PoRoVi, MJor}. The quasi-satellites can be sub-classified in three domains: heliocentric, planetocentric, and `limbo', for which both the Sun and the planet are important and none of them can be treated as a perturbation \citep{2017PoRoVi, Voya}. In this study, we investigate heliocentric quasi-satellite dynamics, although the data analyzed were derived from trajectories residing in the transitional (`limbo') regime. Quasi-satellite orbits have been studied in the context of the motion of asteroids and satellites \citep[\eg][]{2004Mk,2007KN,Wajer2010,2012DD} and for mission design \citep[\eg][]{Naoko, MJor, TASTE, Henon2025}. \citet{2017PoRoVi} provides a comprehensive review of quasi-satellite orbits and their treatment. 

Horseshoe trajectories were first considered by \citet{1911Br} in an approximation of the planar CR3B problem. Later, some orbits and families of orbits of this kind were numerically identified in the Sun--Jupiter system \citep{1961Ra}, followed by numerous others in the general three-body problem \citep[\eg][]{2001LlOl,2006BaOl,2013BeFaPe}. Several analytical theories have been developed to describe the long-term dynamics of horseshoe orbits in the three-body problem, mainly in the case of the Janus--Epimetheus co-orbital pair \citep[\eg][]{1981DeMu,1983YoSyYo,1985SpWa,2013RoPo,2019CoPaYa}.
        
Objects in co-orbital motion can experience transitions between regimes due to the orbits having non-negligible inclinations or to $n$-body interactions \citep[\cf][]{namouni99,Christou00}.  Earth's co-orbitals are also the proximal source region of Earth's `minimoons' as they are the only objects that can energetically be captured in the Earth-Moon system \citep{Granvik2012-minimoons}.

Among the known co-orbitals, \Kamo is particularly noteworthy. With an absolute magnitude\footnote{\url{https://ssd.jpl.nasa.gov/tools/sbdb_lookup.html\#/?sstr=469219}\label{JPL_Kamo}} of $H=24.3$, it is estimated to have a diameter\footref{JPL_Kamo} in the range from $24\meter$ to $107\meter$, and its orbit undergoes transitions between the horseshoe and quasi-satellite regimes.  Spectroscopic studies \citep{Sharkey2021-2016HO3} indicate that it may have a lunar origin which has motivated a series of dynamical studies to evaluate the possibility that lunar ejecta become co-orbitals \citep{CastroCisneros2023,Jiao2024-Kamooalewa,Sfair2025,Zhu2025}. Indeed, \citet{Jiao2024-Kamooalewa} suggest that they have traced its origin to the Giordano Bruno crater, a $1-10\Myr$ old and $\sim22\km$ diameter crater \citep[\eg][]{Morota2009}.  That origin story was subsequently studied by \citet{Fenucci2026} who claim that it is unlikely that \Kamo\ was ejected from Giordano Bruno. In any event, this hypothesis will be tested \emph{in-situ} by the Tianwen-2 spacecraft which was launched in May 2025 and will arrive at \Kamo in the summer of 2026 \citep{Zhang2021}.  

Two other Earth co-orbitals have also been spectroscopically associated with lunar ejecta: the decameter-scale \PT\ \citep{Bolin2025-PT5, Kareta2025-PT5} and the meter-scale \CD\ \citep{Bolin2020-CD3}.  Both objects had a negative binding energy with Earth but only \CD\ satisfied the conditions to be informally classified as a `minimoon' \citep{Fedorets2017-minimoons}.  Intriguingly, backwards propagation of \CD's trajectory suggests that it had a 1.4\% chance of intersecting the lunar surface in September of 2017, at the time it was captured into its geocentric orbit.  This suggests that it might have been ejected by an impact at that time but \cite{Fedorets2020-2020CD3} rejected that argument for a number of reasons.  Even so, it may have been ejected in an impact long ago and remained in, or near, Earth co-orbital space to be captured again in 2017.

While it has long been known that lunar ejecta \emph{can} evolve onto Earth-like orbits, how often does it happen and what fraction of the steady-state co-orbital population has a lunar provenance?  For instance, the near-Earth object model NEOMOD3 predicts that there are \nCoorbitalsLikeKamo\ objects of main belt origin with an orbit and absolute magnitude similar to \Kamo, \ie\ with $H<27$, $0.05\leq e \leq0.15$, and $5\arcdeg \leq i \leq 15\arcdeg$. The largest known Earth co-orbital is the $\sim\fLargestDiameterMeters\meter$ diameter\footnote{Assuming an albedo of $0.143$ that assigns an $H=17.75$ object a diameter of $1\km$.\label{ftn.albedo}} (3753) Cruithne and it seems unlikely that an object of that size could be liberated from the Moon's surface by an asteroid impact with sufficient speed to escape the Earth-Moon system, transition into an Earth-like orbit, and remain there long enough to be identified in a co-orbital state at this time, given that the dynamical lifetime of objects in near-Earth space is on the order of millions of years \citep[\eg][]{Gladman1997-DynamicalLifetimesOfObjectsInjectedIntoAsteroidBeltResonances}.

In addition to the scientific merits of understanding the provenance of objects in the inner solar system, the Earth's co-orbitals are some of the lowest $\deltav$ mission targets \citep[\eg][]{Granvik2013,Jedicke2018-broken-plane}.  They are thus interesting targets for space exploration as the solar system's dynamics have performed the task of transporting objects from the Moon and main belt into more easily accessible orbits.  Their economic and strategic viability will also depend on their mineralogy which is different for the two sources.

%##################################################
\section{Method}
\label{s.Method}

%---------------------------------------------
\subsection{Co-orbital dynamics}
\label{ss.Method_co-orbital_dynamics}

The integrations for this work were performed by \citet{Jedicke2025} who calculated the steady-state size-frequency and orbital element distribution of minimoons generated by lunar impacts.  We used the results of their integrations during the time periods in which the particles were in heliocentric orbit.

We refer the reader to \citet{Jedicke2025} for the details of their simulation and only provide a high level summary here.  Their work simulated the fate of 12,000 particles launched from 100 random locations on the Moon's surface at 19 different speeds, 6 azimuth angles, and $45\arcdeg$ elevation angle.  The particles were launched at evenly spaced times within a Metonic cycle of $\sim19\yr$ and integrated using REBOUND \citep{ReinLiu2012} with the 10 most massive solar system objects as perturbers.  The particles were integrated for up to $55\Myr$ or until they left a torus enveloping Earth's orbit.

The state of each particle during a heliocentric phase was recorded every 7,500$\Days$, $\sim20.5\yr$, including the heliocentric orbital elements and resonant angle with the ecliptic as the reference plane.  The time step was selected to allow the identification of co-orbital behavior while also considering data storage limitations.  Particles typically remain in the horseshoe and tadpole states for hundreds of years, whereas quasi-satellite states persist only for a few decades \citep[\eg][]{DiRuzza2023, Cortese2025}. 

Earlier studies of the dynamical evolution of ejected lunar material were similar to this work but differed in the details of their assumptions and focus \citep[\eg][]{CastroCisneros2023, Jiao2024-Kamooalewa,Zhu2025,Sfair2025}.  The most recent studies typically employed the REBOUND integrator that was used here but with different sets of perturbers. The propagation intervals and time steps were both shorter and longer than ours.  Some of the studies specifically examined ejecta from the Giordano Bruno crater while others used random locations on the lunar surface.  Our study used the largest number of launch speeds although some of the others considered a wider range.  While the different studies differ in their details the integrations tend to be insensitive to some of the initial conditions.  For instance, \citet{Sfair2025} showed that their results were unaffected by both the altitude and azimuth of the ensemble of particles's launch from the lunar surface and \citet{Zhu2025} state that `the orbital dynamics of escaping ejecta are largely insensitive to the launch positions and angles'.

\begin{figure}
    \centering
    \includegraphics[width=8cm]{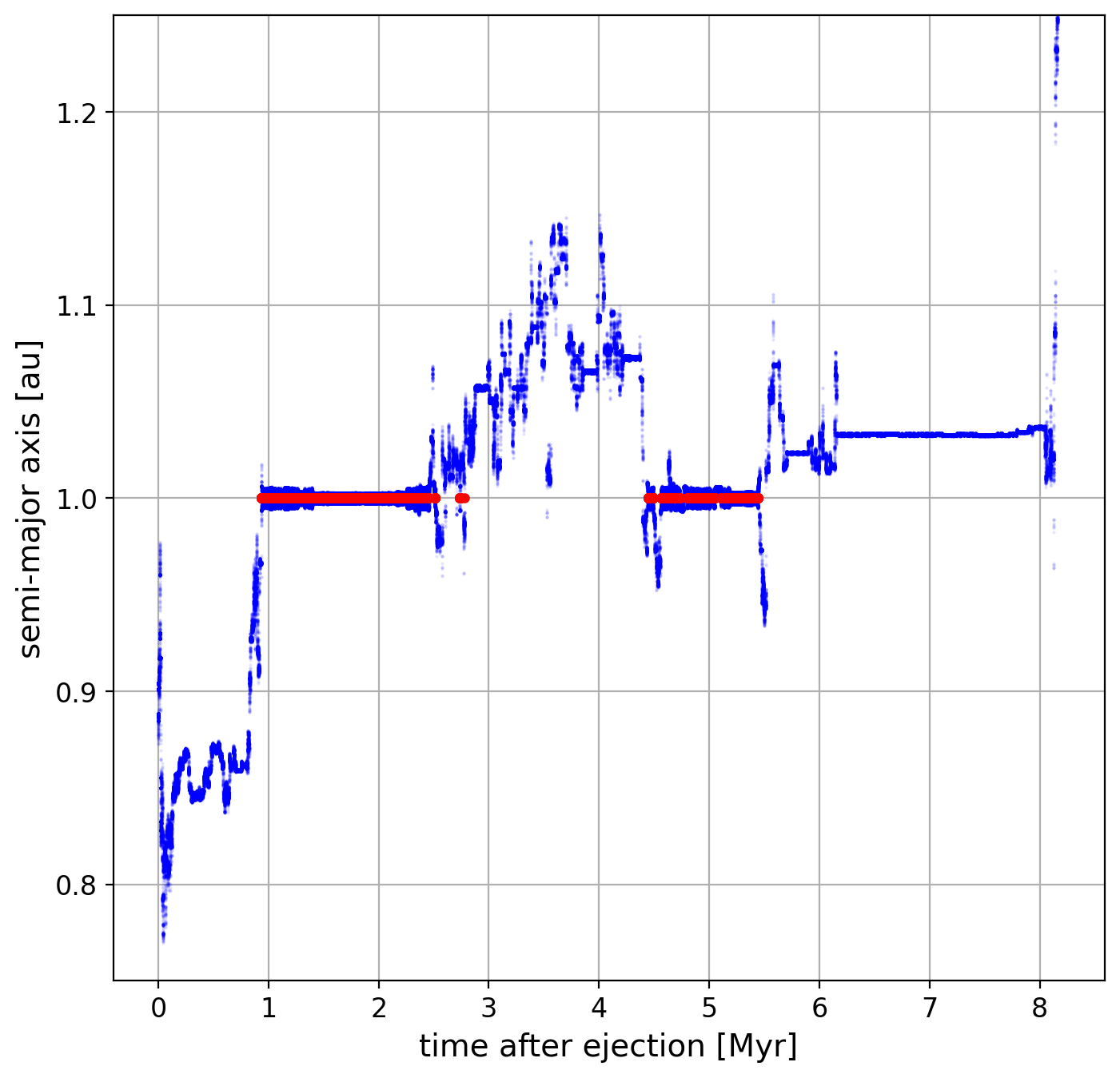} 
    \includegraphics[width=8cm]{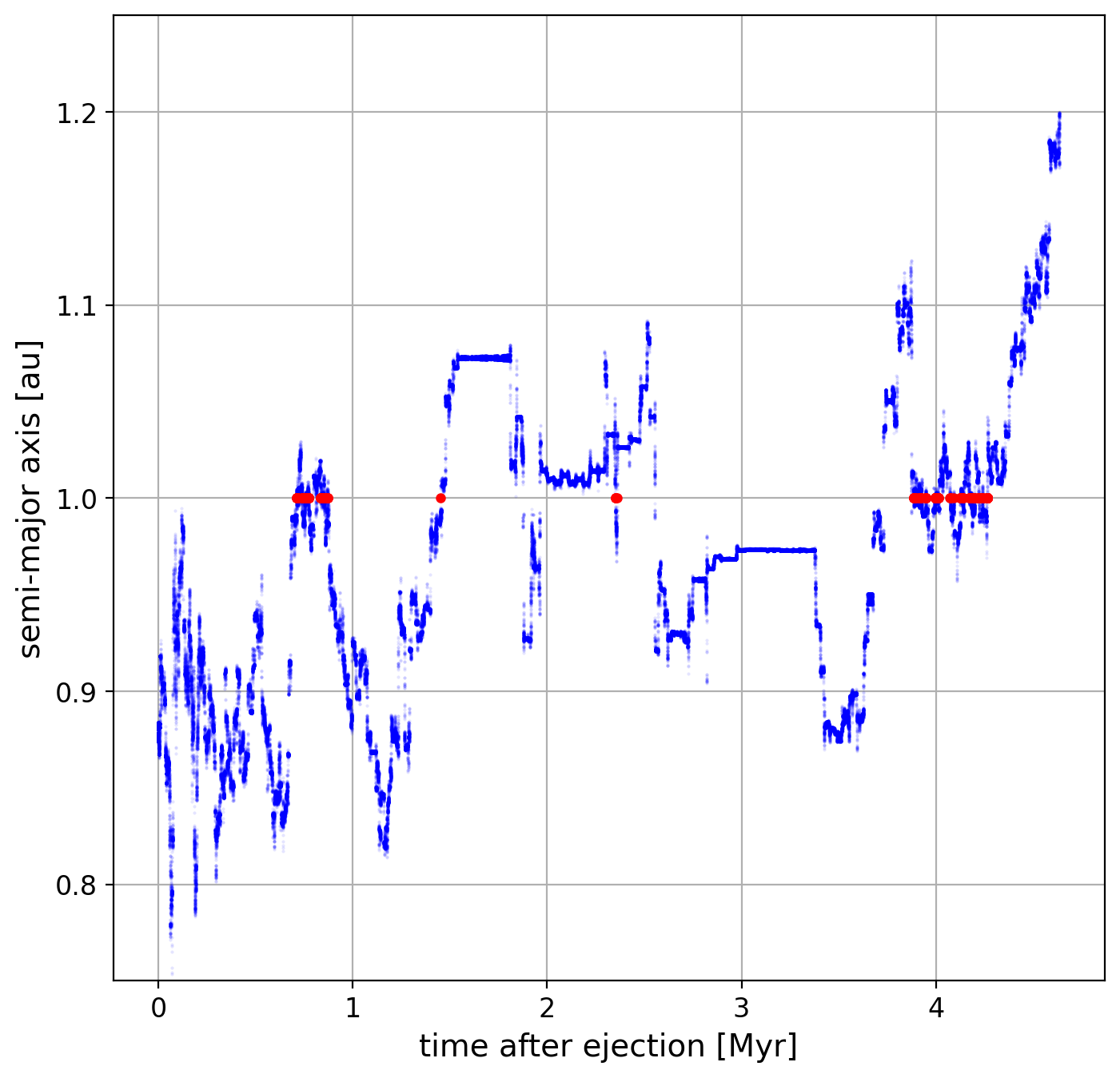}
    \caption{(blue) Semi-major axis time series highlighting (red) candidate co-orbital intervals for two test particles.}
    \label{fig.axis_time_evolution}
\end{figure}

We identified the particles' co-orbital regimes and time periods based on the software developed by \cite{DiRuzza2023} extended with the techniques developed in \cite{Cortese2025}.  The first step was to identify co-orbital candidates by pinpointing times at which a particle's semi-major axis crossed $a=1\au$ and consider the time intervals during which the variation in semi-major axis is $\le3$\%  (\fig{fig.axis_time_evolution}).

The regimes were then distinguished by the behavior of the resonant angle, $\theta$, the difference between the longitudes of the asteroid, $L$, and Earth, $L_E$,
\begin{equation}
    \theta  = L - L_E 
            = (\Omega+\omega+M) - (\Omega_E+\omega_E+M_E),
\end{equation}
where the $E$ subscript refers to the orbital elements of Earth, the parameters without subscripts refer to the particle, $\Omega$ is the longitude of the ascending node, $\omega$ is the argument of pericenter, and $M$ is the mean anomaly.

\begin{figure}
    \centering
    \includegraphics[width=8cm]{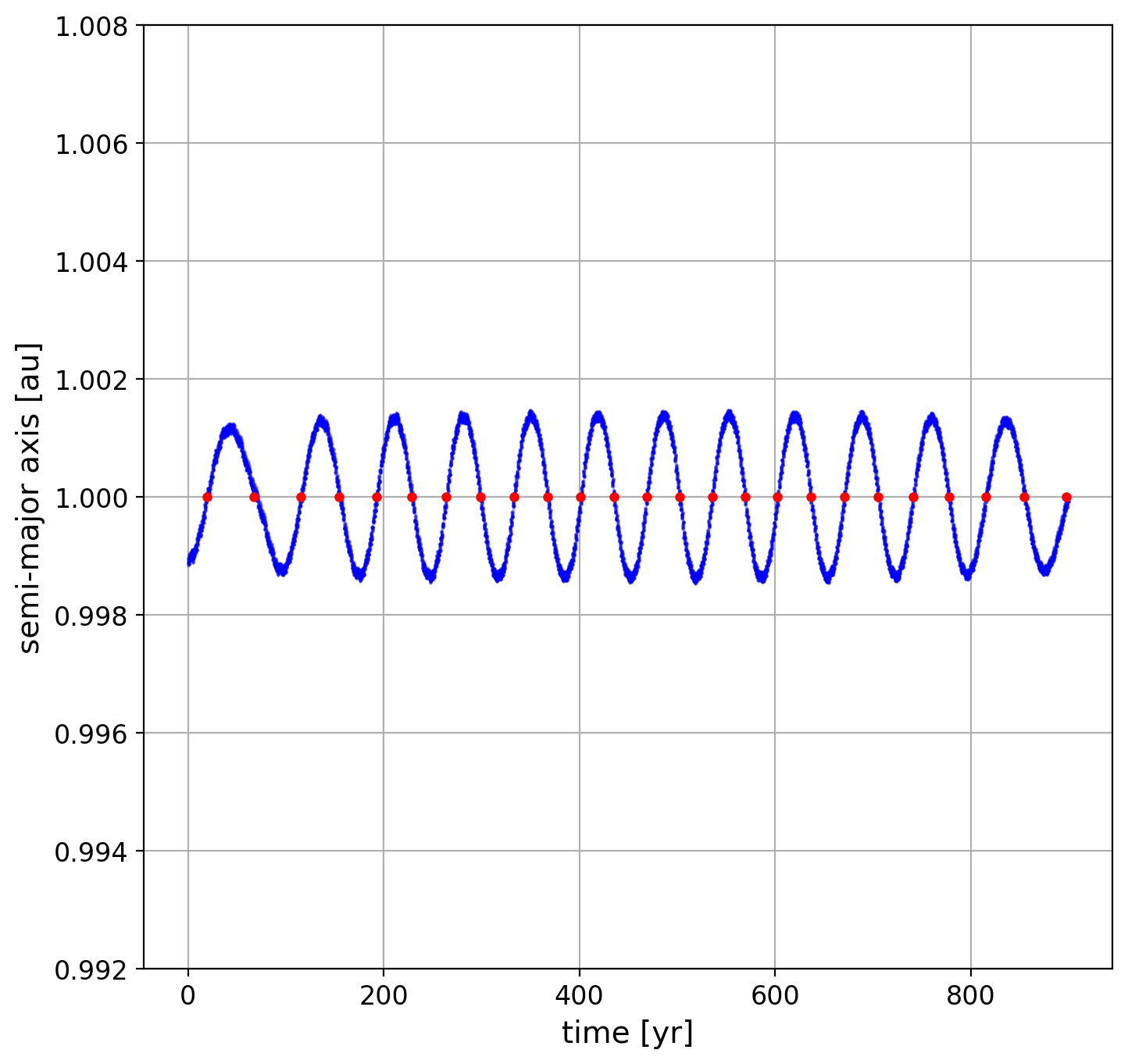}
    \includegraphics[width=8cm]{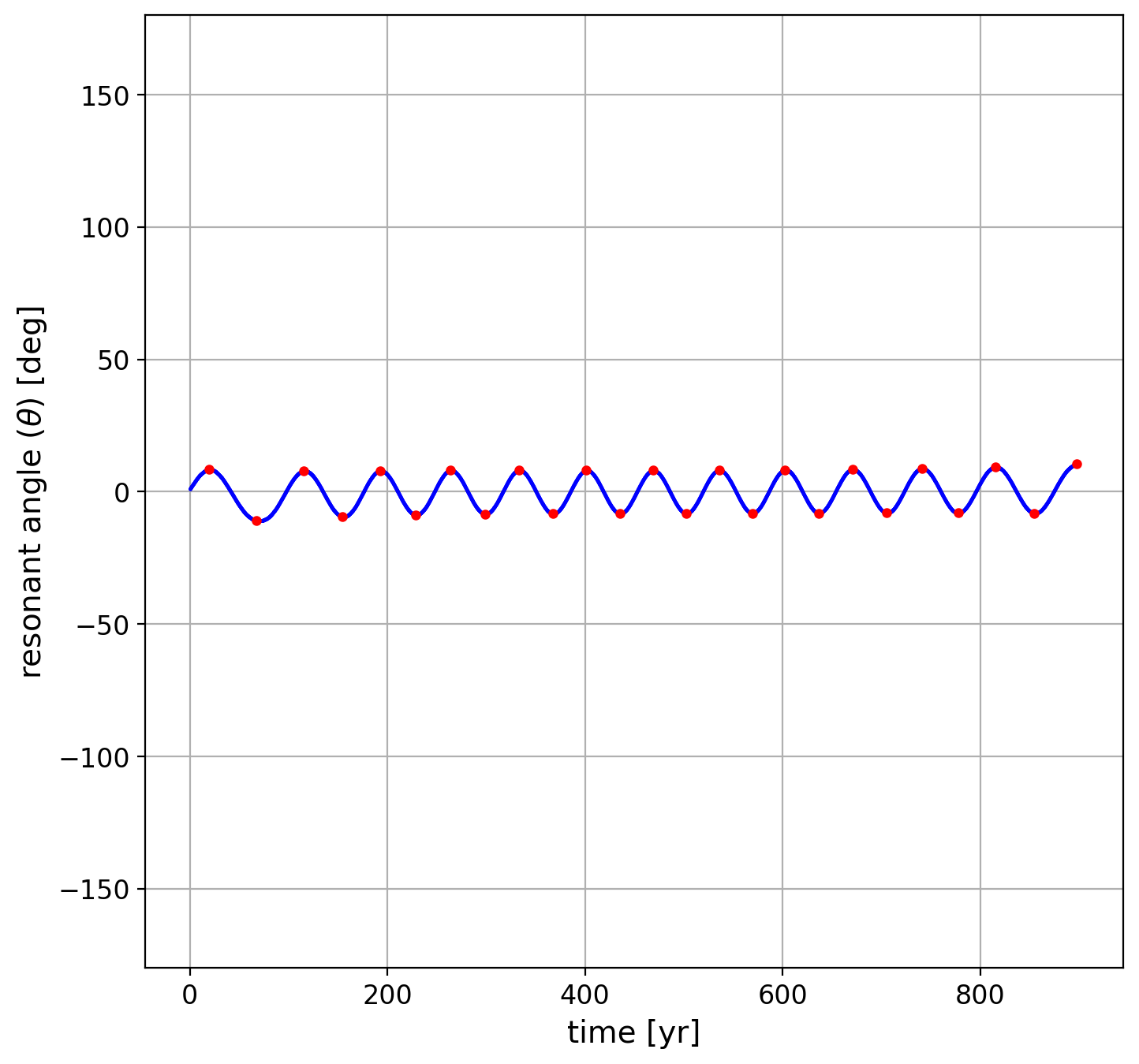}
    \caption{
        The quasi-satellite behavior of 2004~GU$_9$ in ({\bf left}) semi-major axis and ({\bf right}) the resonant angle, $\theta$. The red points correspond to the  $a=1\au$ crossings.  The $y$-ranges in both panels of Figs~\ref{fig.quasi-satellite}-\ref{fig.Kamo} are the same to emphasize the differences between them.}
    \label{fig.quasi-satellite}
\end{figure}

\begin{figure}
    \centering
    \includegraphics[width=8cm]{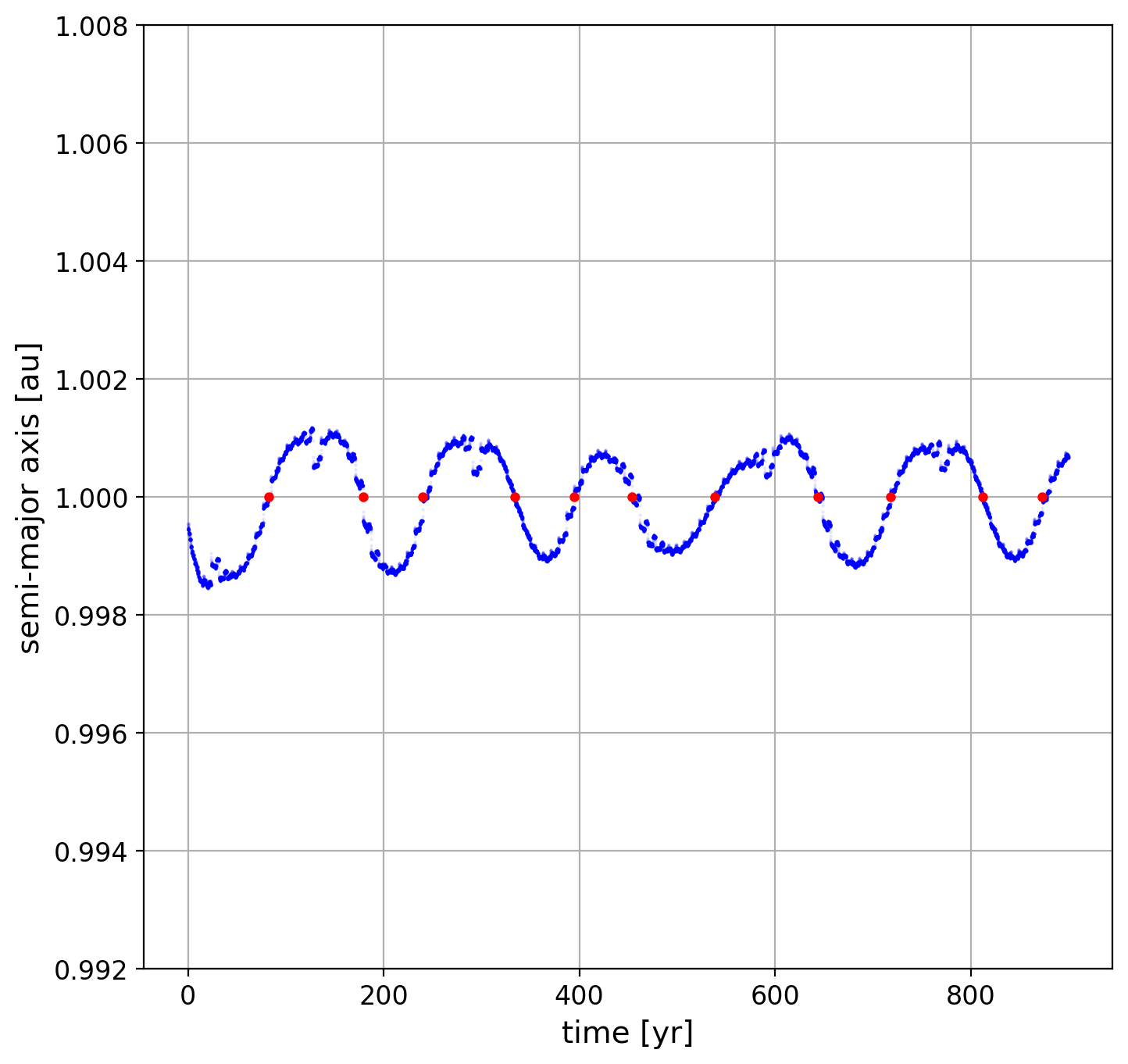} \includegraphics[width=8cm]{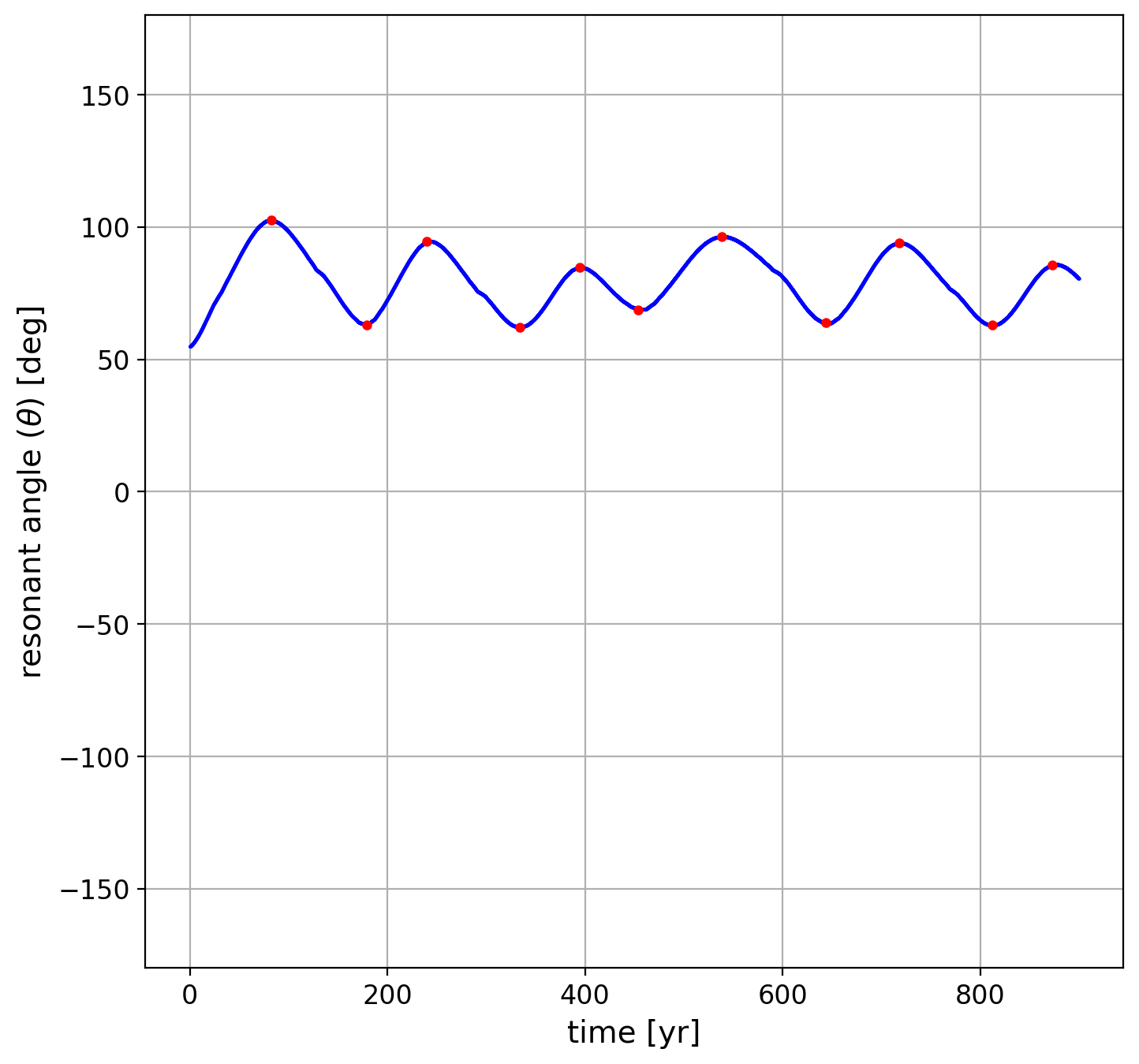}
    \caption{
        The tadpole behavior of 2020~XL$_5$ in ({\bf left}) semi-major axis and ({\bf right}) the resonant angle, $\theta$. The red points correspond to the $a=1\au$ crossings.}
    \label{fig.tadpole}
\end{figure}

\begin{figure}
    \centering
    \includegraphics[width=8cm]{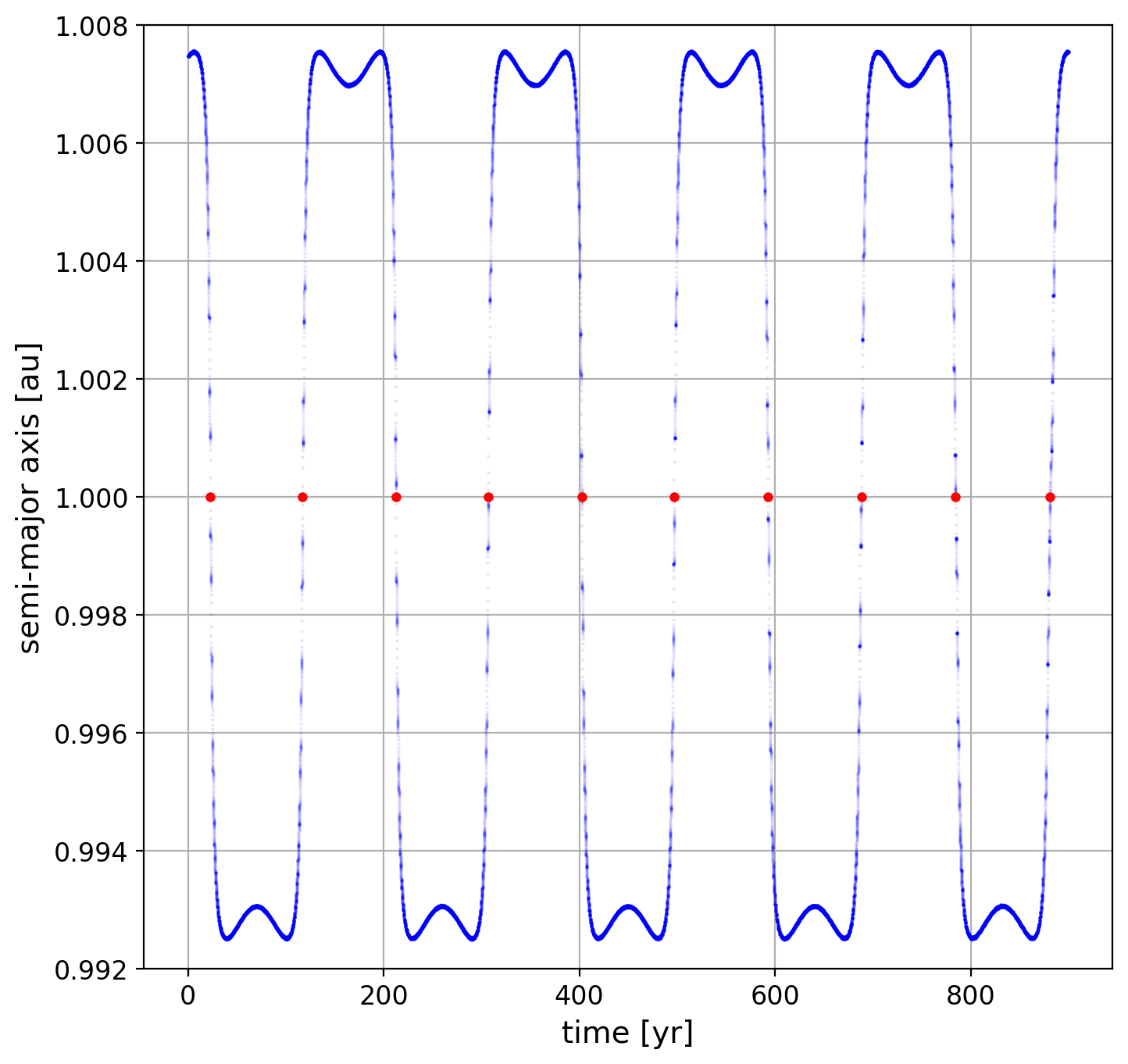} \includegraphics[width=8cm]{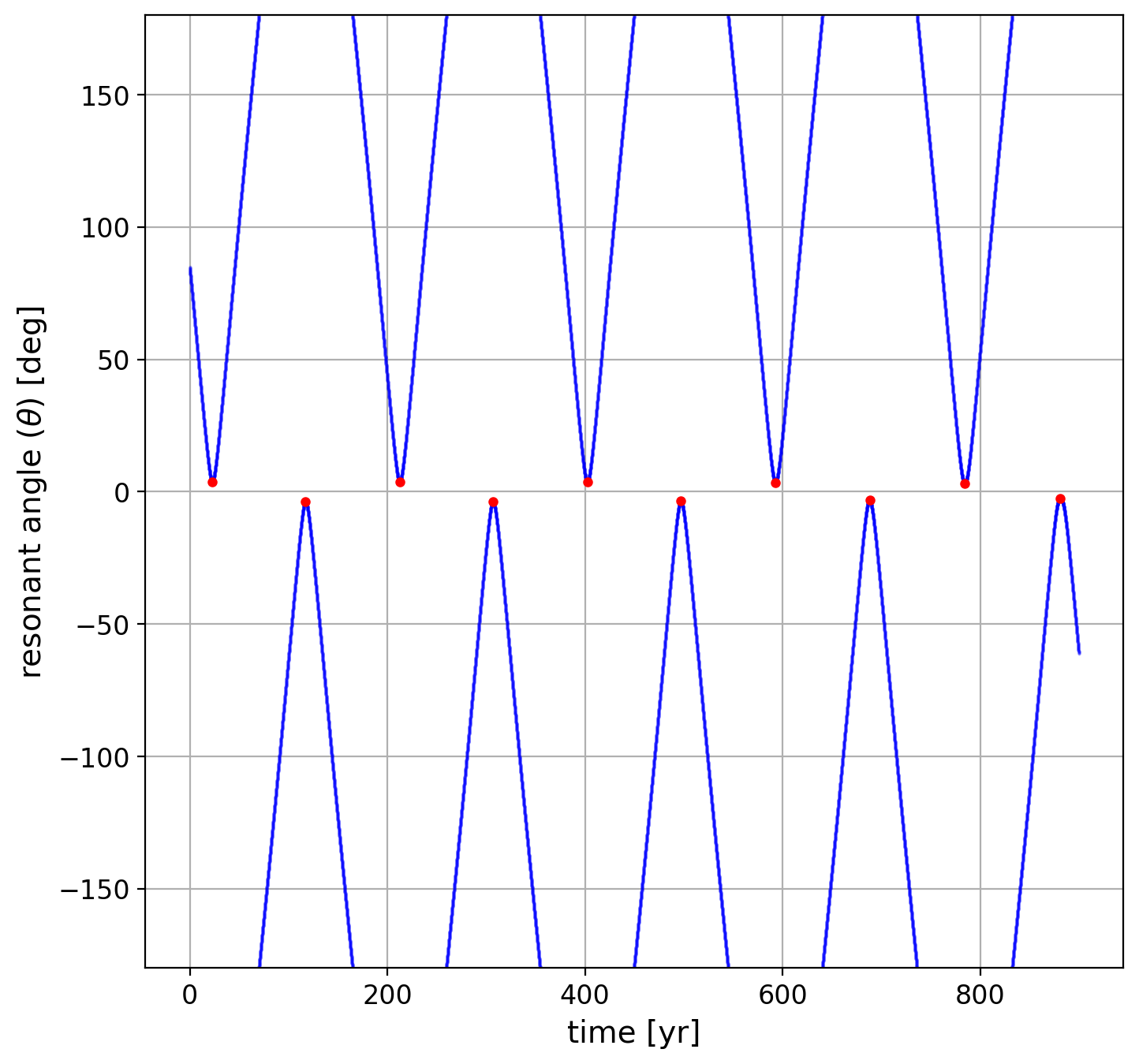}
    \caption{
        The horseshoe behavior of 2002~AA$_{29}$ in ({\bf left}) semi-major axis and ({\bf right}) the resonant angle, $\theta$. The red points correspond to the $a=1\au$ crossings.}
    \label{fig.horseshoe}
\end{figure}

\begin{figure}
    \centering
    \includegraphics[width=8cm]{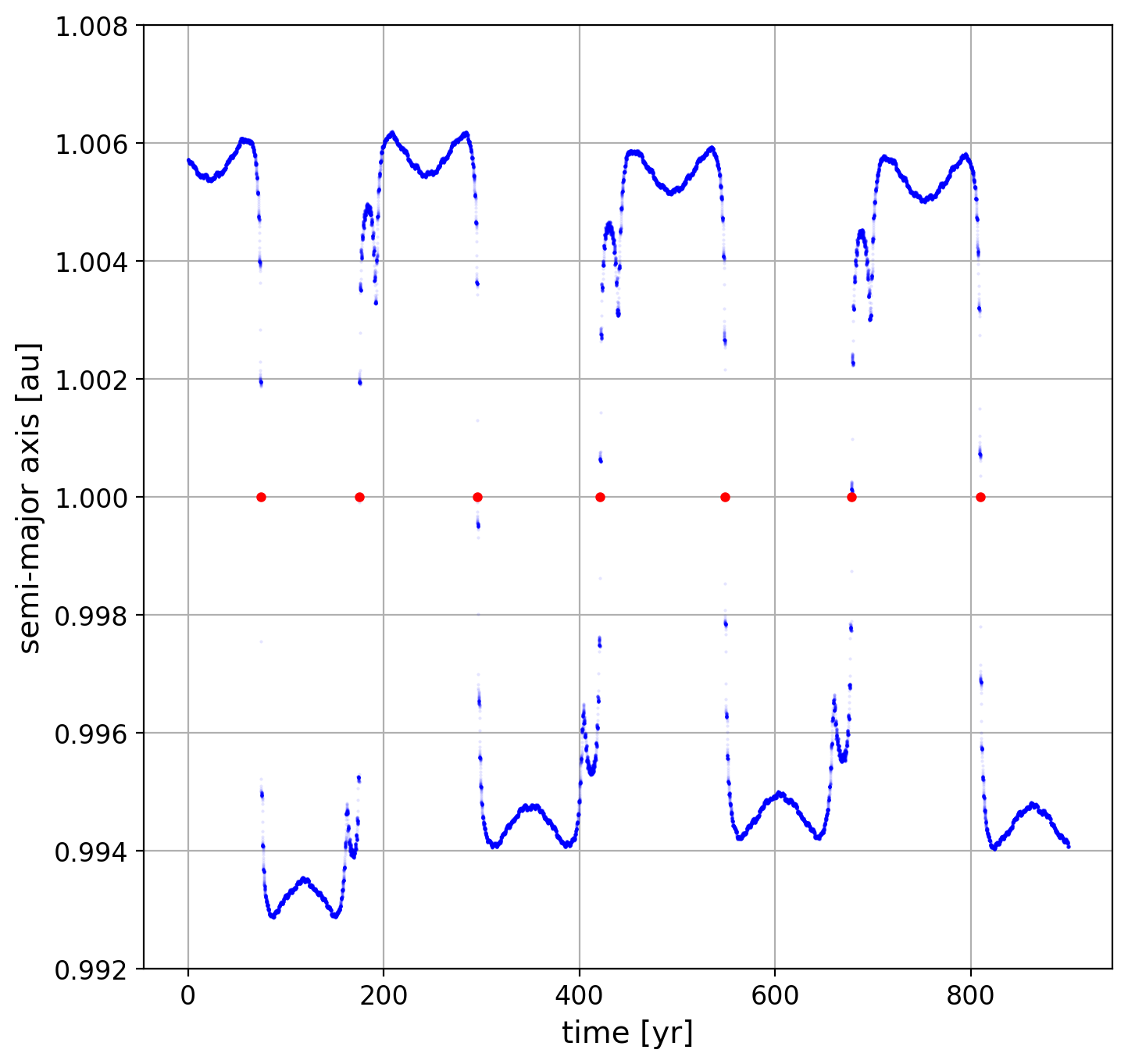} 
    \includegraphics[width=8cm]{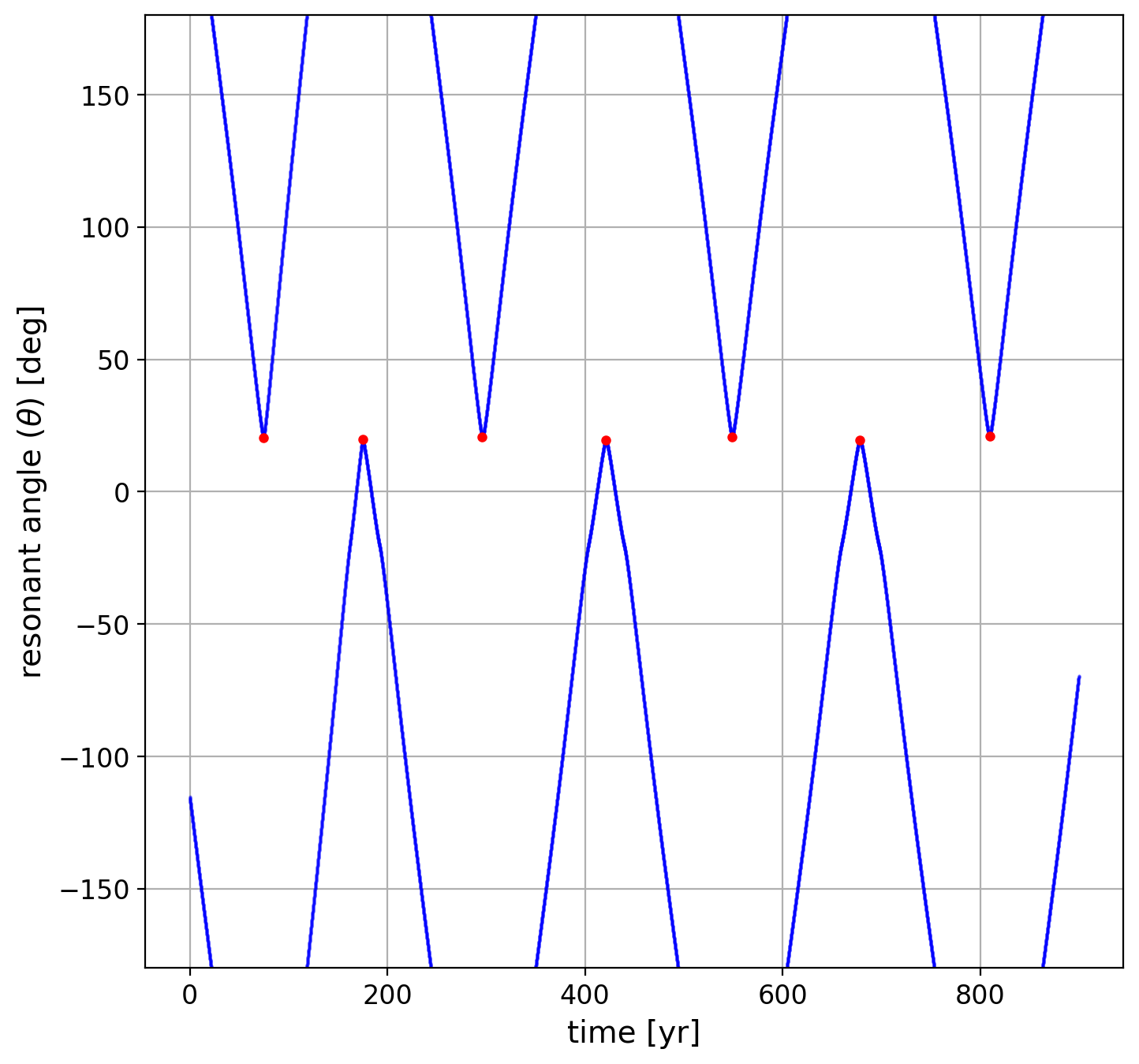}
    \caption{
        The compound behavior of 2020~CX$_1$ in ({\bf left}) semi-major axis and ({\bf right}) the resonant angle, $\theta$. The red points correspond to the $a=1\au$ crossings.}
    \label{fig.CP}
\end{figure}

\begin{figure}
    \centering
    \includegraphics[width=8cm]{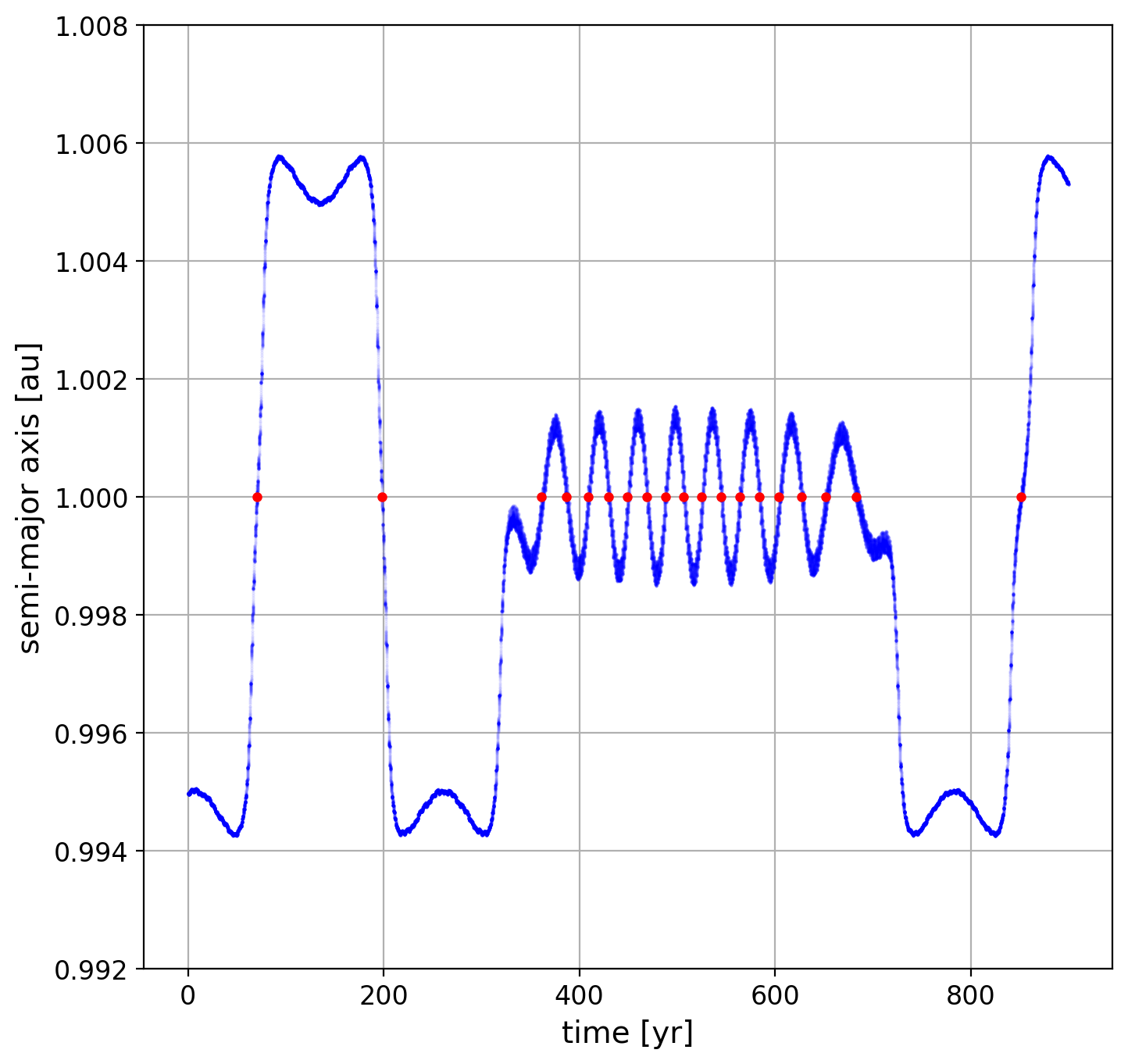} 
    \includegraphics[width=8cm]{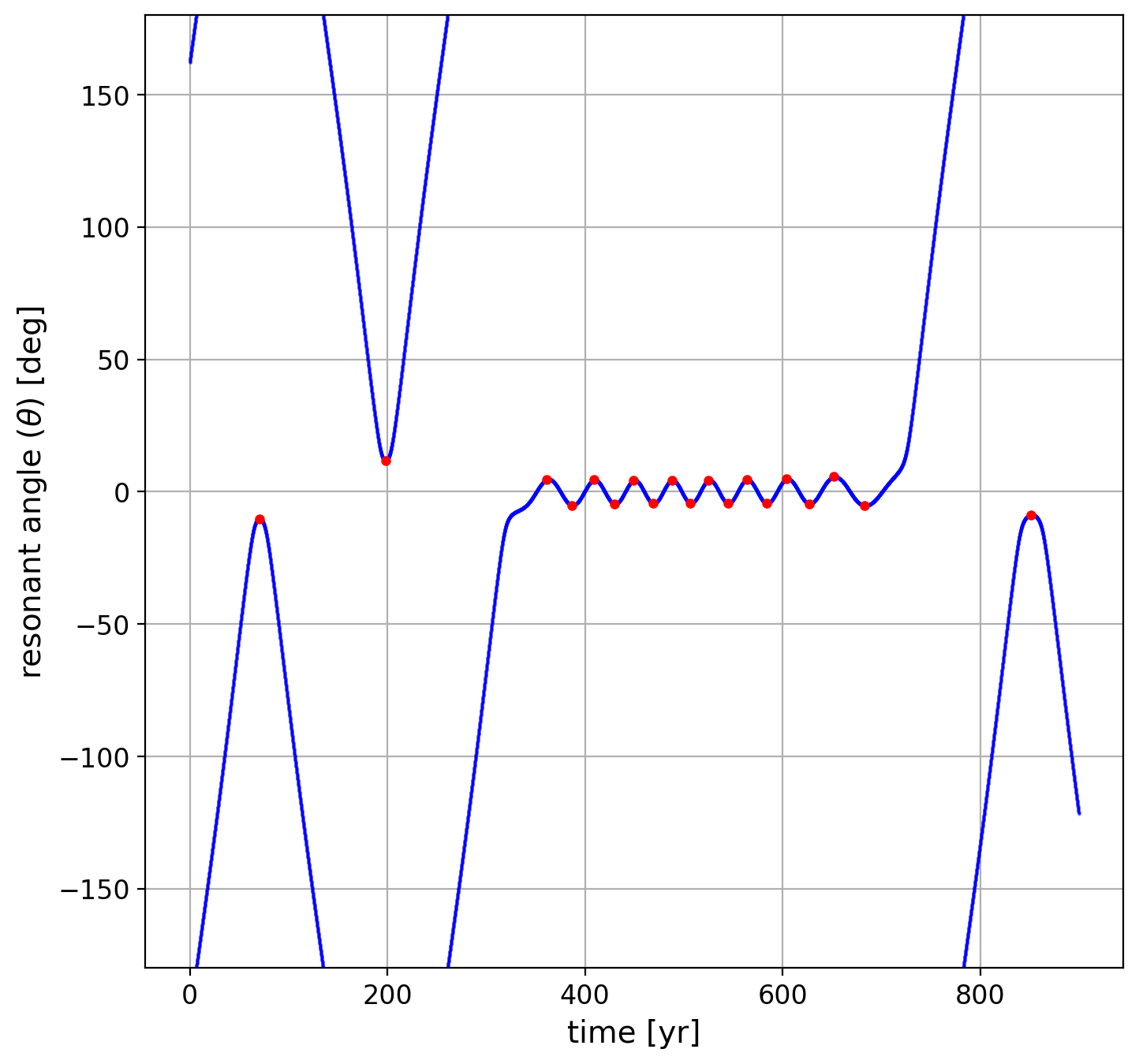}
    \caption{
        The transition between horseshoe and quasi-satellite motions experienced by \Kamo\ in ({\bf left}) semi-major axis and ({\bf right}) the resonant angle, $\theta$. The red points correspond to the $a=1\au$ crossings.}
    \label{fig.Kamo}
\end{figure}

Following \citet{Pousse+Alessi2022} and \citet{Cortese2025} the co-orbital regimes are formally defined as
\begin{itemize}

    \setlength\itemsep{1pt}   % space between items

    \item \emph{Quasi-satellite} (QS):
    when $\theta$ oscillates around $0^{\circ}$ (\fig{fig.quasi-satellite});

    \item \emph{Tadpole} (TP):
    when $\theta$ oscillates around $\pm 60^{\circ}$  (\fig{fig.tadpole});

    \item \emph{Horseshoe} (HS):
    when $\theta$ oscillates around $180^{\circ}$ (\fig{fig.horseshoe});

    \item \emph{Compound} (CP):
    when $\theta$ oscillates around $180^{\circ}$ but, as a combination of the horseshoe and quasi-satellite regimes, $\theta$ crosses $0^{\circ}$ (\fig{fig.CP}). The CP regime is different from a transition between a horseshoe and a quasi-satellite regime as experienced by Kamo`oalewa (\fig{fig.Kamo}).
    
\end{itemize}

The algorithm we employed identifies the co-orbital regime by determining the epoch at which $a=1\au$ (\eg\ the red points in the left panels of figs.~\ref{fig.quasi-satellite}-\ref{fig.CP}) at which the resonant angle is at a local minimum or maximum, $\theta_{min}$ and $\theta_{max}$ respectively (\eg\ the red points in the right panels of figs.~\ref{fig.quasi-satellite}-\ref{fig.CP} where $\dif^2 \theta/\dif t^2>0$ and $<0$ respectively). With this definition it is possible that $\theta_{max}<\theta_{min}$. The particles are then associated with a co-orbital regime according to the following criteria:

\begin{itemize}

    \item {\it Quasi-satellite (QS)}:
    when ($\theta_{max}\times\theta_{min} < 0$ and $\theta_{max} > \theta_{min}$) or ($\theta_{max}\times\theta_{min} > 0$ and $|\theta_{max} + \theta_{min}|/2 < 35^{\circ}$, a `drifting' quasi-satellite) ;
    
    \item {\it Tadpole (TP)}:
    when $\theta_{max}\times\theta_{min} > 0$ and $\theta_{max} > \theta_{min}$ and $|\theta_{max} + \theta_{min}|/2\geq 35^{\circ}$;
    
    \item {\it Horseshoe (HS)}:
    when $\theta_{max}\times\theta_{min} < 0$ and $\theta_{max} < \theta_{min}$;
    
    \item {\it Compound (CP)}:
    when $\theta_{max}\times\theta_{min} > 0$ and $\theta_{max} < \theta_{min}$.
    
\end{itemize}

\begin{figure}
    \centering
    \includegraphics[width=8.15cm]{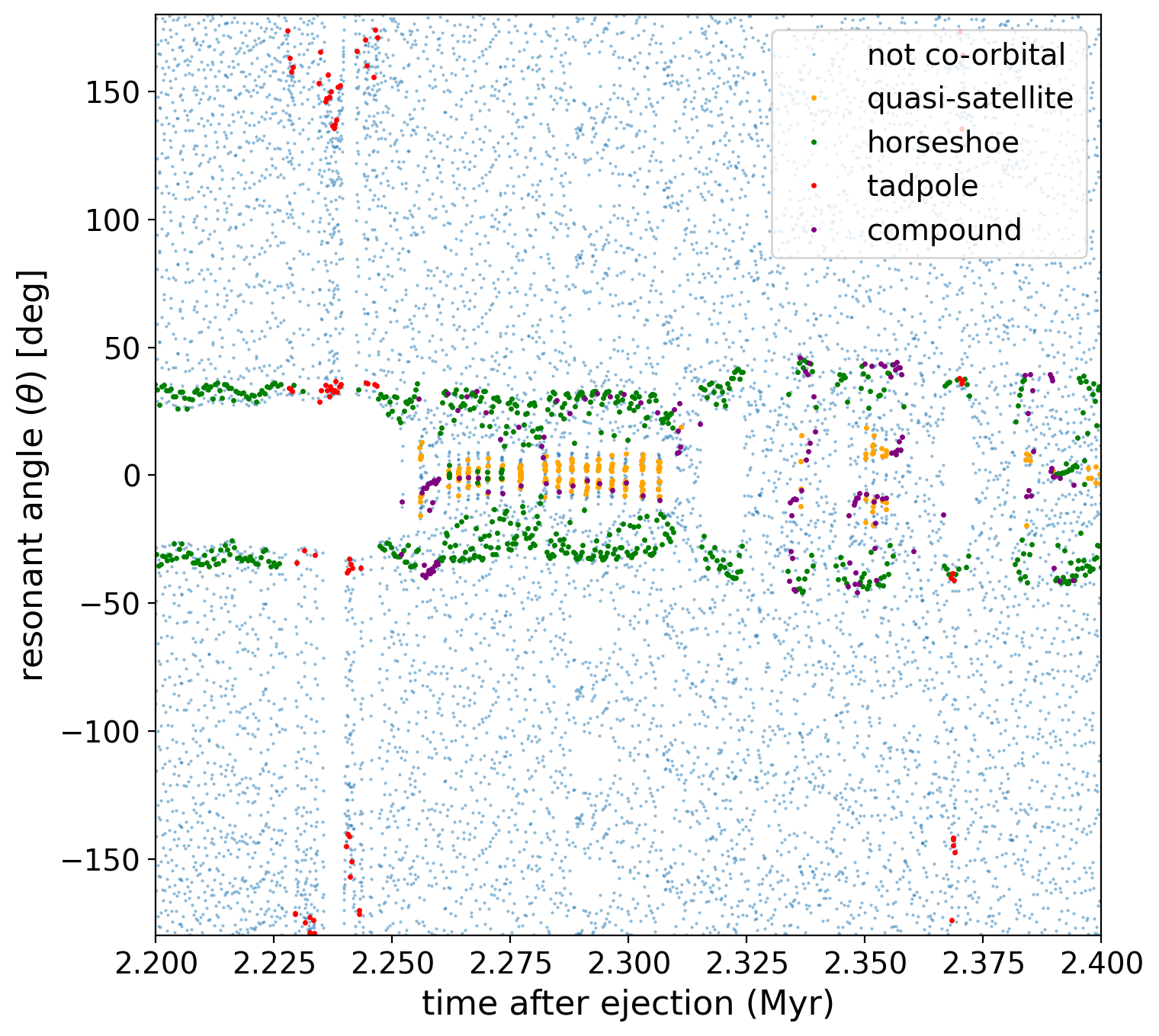} 
    \includegraphics[width=7.85cm]{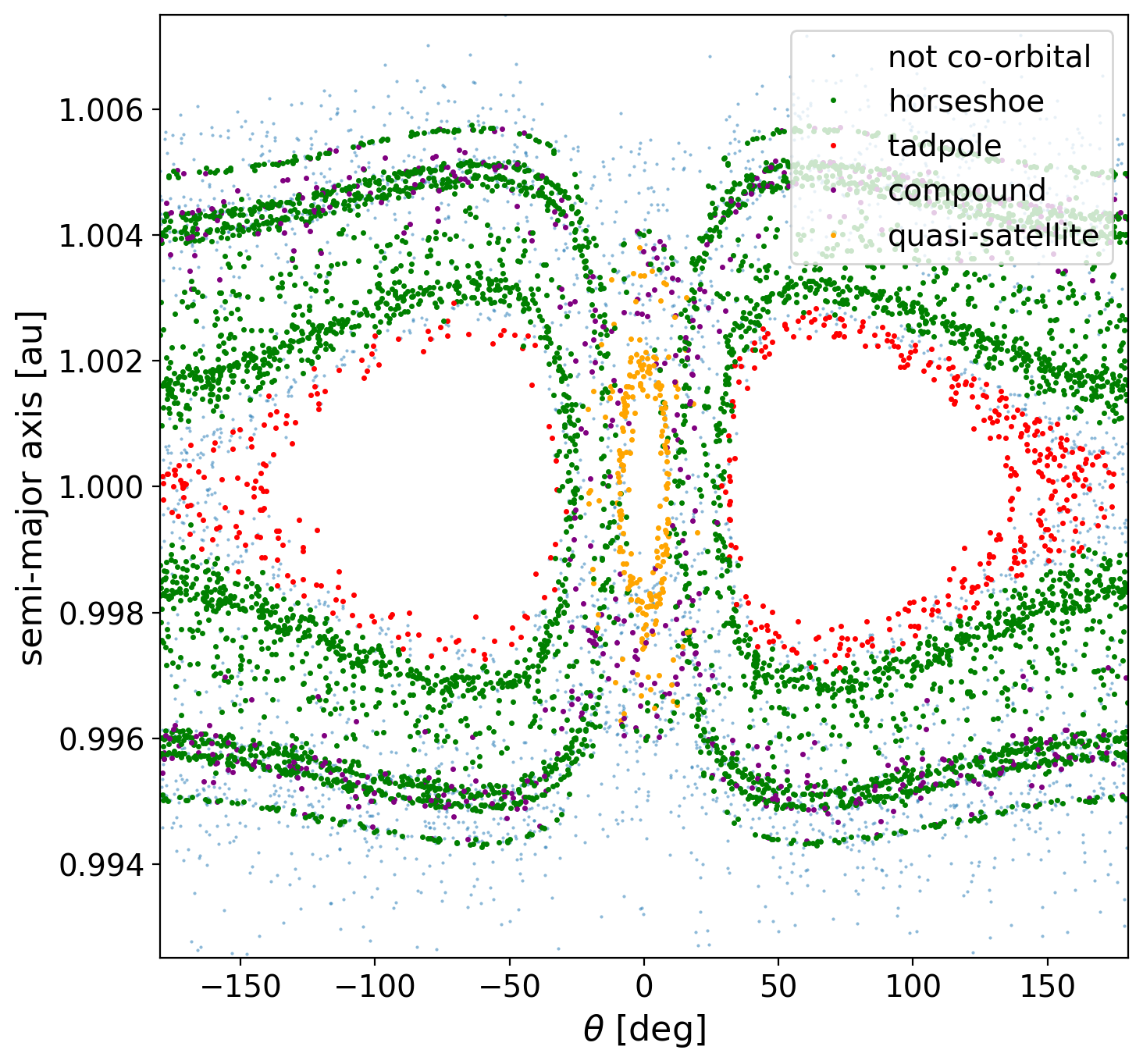}
    \caption{
        ({\bf left}) The time evolution of the resonant angle, $\theta$, over 200~kyr for one of our synthetic lunar ejecta particles indicating times when it is not co-orbital and in a co-orbital regime. ({\bf right}) The behavior of the same particle in $(\theta,a)$ phase space illustrating well-defined structures corresponding to the different co-orbital regimes. This example corresponds to part of the trajectory shown on the left panel of \fig{fig.axis_time_evolution}.}
    \label{fig.theta_and_semiaxis_time_evolution}
\end{figure}

\begin{figure}
    \centering
    \includegraphics[width=0.32\columnwidth]{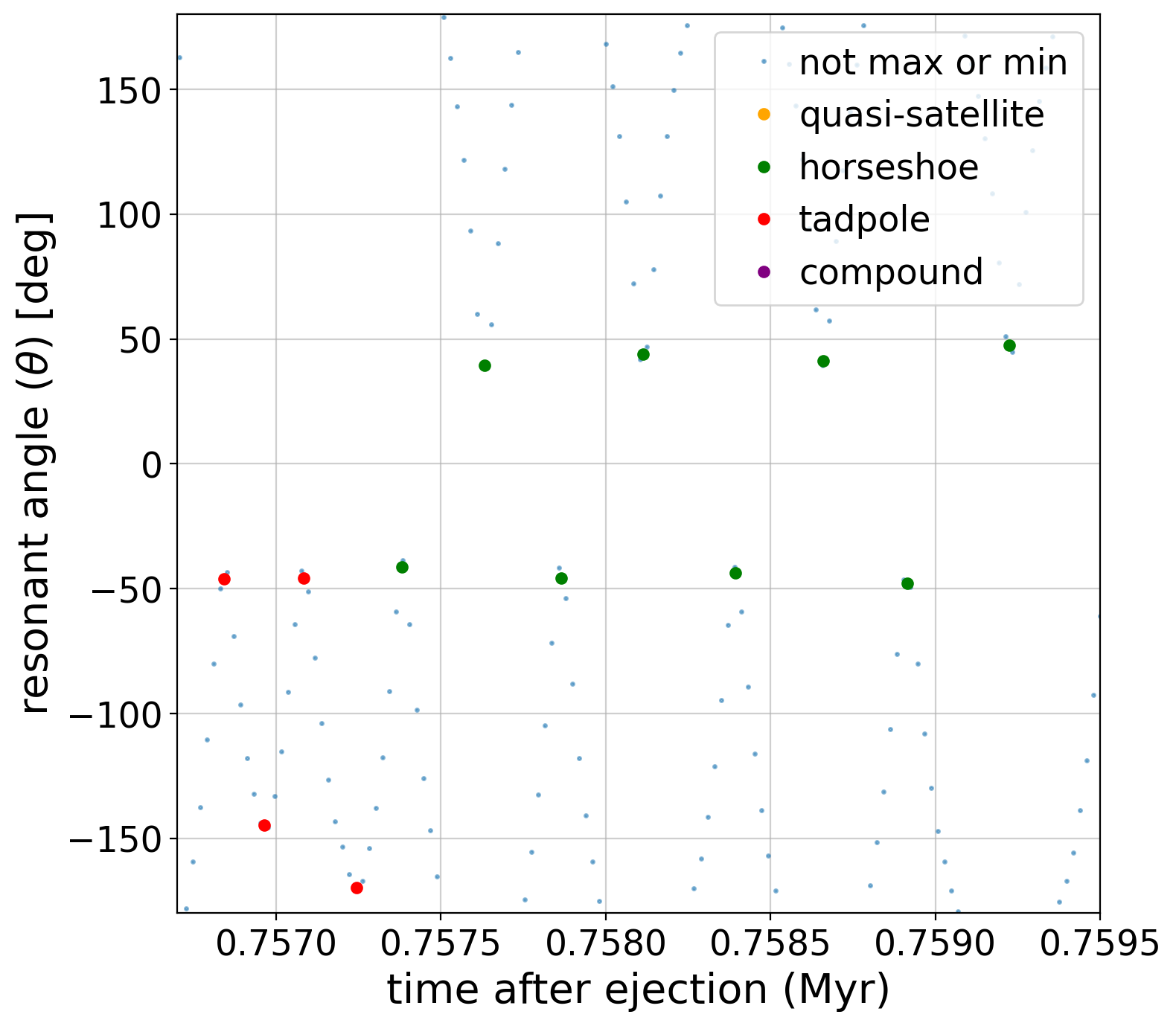} 
    \includegraphics[width=0.305\columnwidth]{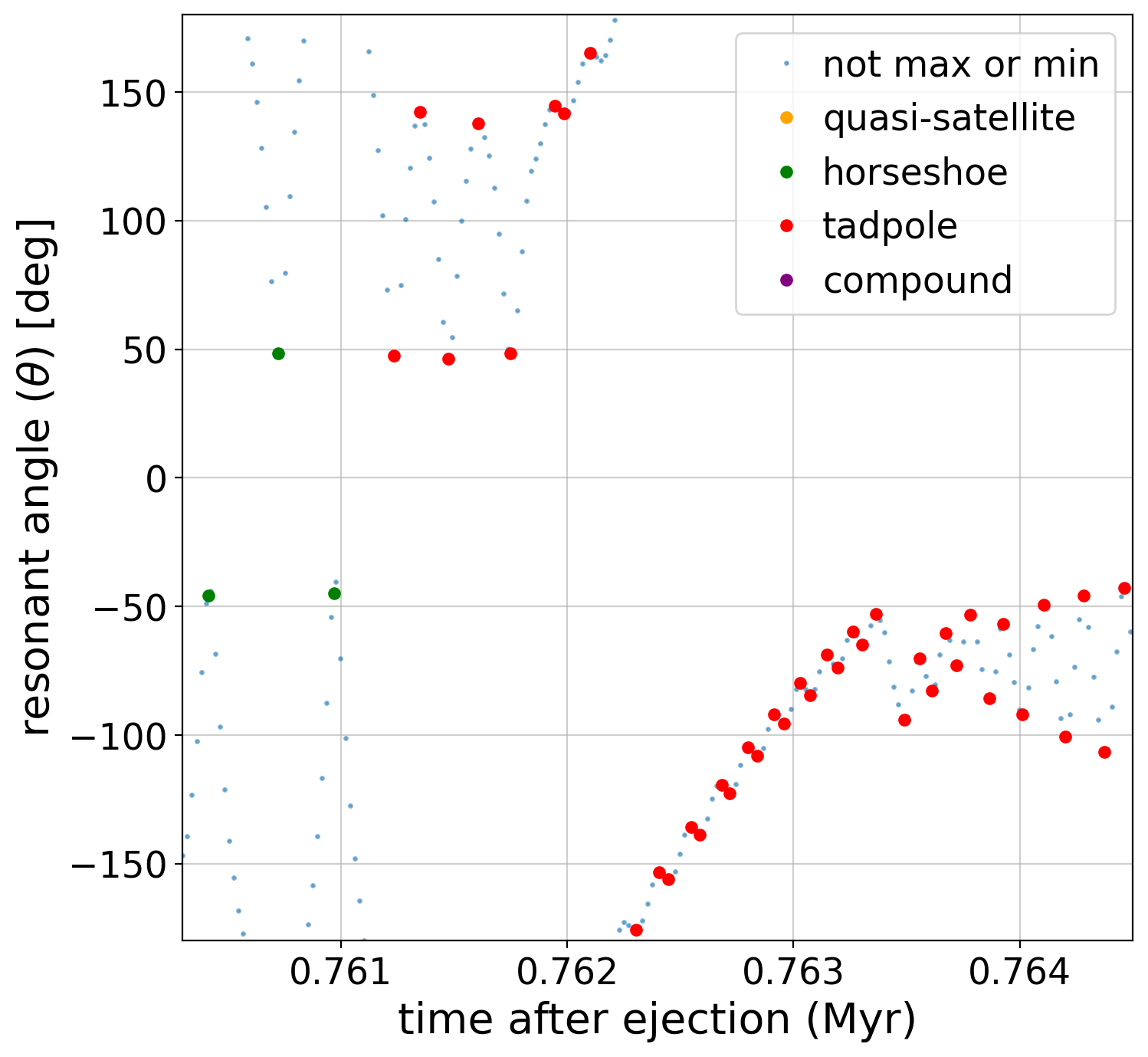}
    \includegraphics[width=0.32\columnwidth]{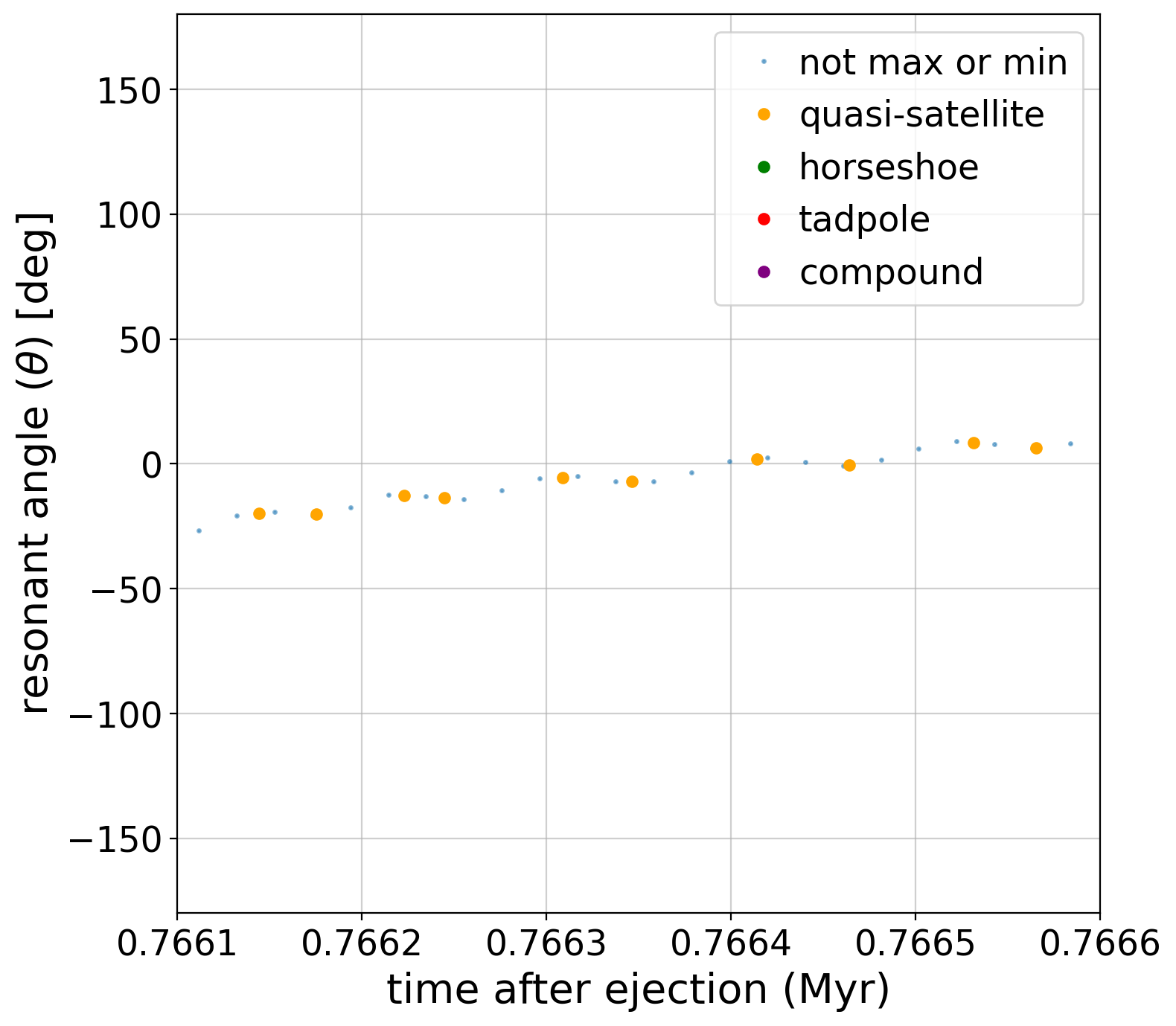}
    \caption{High time resolution behavior of the resonant angle for the synthetic object on the right panel of \fig{fig.axis_time_evolution}\ illustrating transitions between co-orbital regimes and the difference between the QS, TP and HS orbits. ({\bf left}) A transition from the tadpole to horseshoe regimes. ({\bf center}) A transition from the horseshoe to tadpole regime. The particle exhibits a `drifting' resonance angle while in the tadpole regime for times $\gtrsim0.7622\Myr$. ({\bf right}) A drifting quasi-satellite regime.}
    \label{fig.highres_theta_time_evolution}
\end{figure}

Known objects in the QS regime (\tab{tab.Earth_co-orbitals_known_1600_2500_NOW}) experience an $a=1\au$ crossing (Fig.~\ref{fig.quasi-satellite}) at least every 20 years, \eg\ \Kamo, and we required co-orbital interval durations to be $\ge100\yr$, \ie\ 5 time steps.
The behavior of the resonant angle while in a co-orbital regime can be irregular due to $n$-body perturbations (figs.~\ref{fig.quasi-satellite}-\ref{fig.Kamo})
which is why the conditions on $\theta$  are more complex than might be expected if we were restricted to the 3-body problem. They also allow for `drifting', secular changes in the average value of the resonant angle, in the quasi-satellite and tadpole regimes, \eg\ the middle and right panels of \fig{fig.highres_theta_time_evolution}, the distinction is that the angle crosses zero for drifting QS, whereas it does not for drifting TP. The value of $35\arcdeg$ was selected by examination of the $(\theta,e)$ map in \cite{Pousse+Alessi2022} to classify quasi-coplanar co-orbital motion. Moreover, tadpole orbits that stem from $L_4/L_5$ are not necessarily centered at $\theta=\pm60\arcdeg$ because this value depends on the orbit's eccentricity \citep{2017PoRoVi}. Finally, note that particles may exhibit more than one co-orbital behavior in their lifetime and that transitions between regimes may occur with no temporal gap at the resolution of our integrations (\fig{fig.theta_and_semiaxis_time_evolution}~ and \fig{fig.highres_theta_time_evolution}). 

The co-orbital identification algorithms employed by others are also based on an initial filtering on the semi-major axis followed by a classification of the co-orbital regime based on the oscillation of the resonant angle \citep[\eg][]{CastroCisneros2023,Jiao2024-Kamooalewa,Sfair2025}. \citet{CastroCisneros2023} state that their final classification was by visual inspection and their results do not show tadpole behavior while \citet{Sfair2025} implemented an automated methodology.

%---------------------------------------------
\subsection{Co-orbital steady-state populations}
\label{ss.Method_co-orbital_steady_state_SFD}

Our calculation of the steady-state population of co-orbitals is essentially identical to that developed by \citet{Jedicke2025} for the minimoon population and we refer the reader to that work for the details.  It derives from the basic steady-state equation:
\begin{equation}
    \bar n(d) = \bar f(d) \; \bar\ell(d).
    \label{eqn.nSteadyState-basic}
\end{equation}
\noindent where $d$ is an object's diameter, $\bar n$ is the average number of objects in the steady-state, $\bar f$ is the flux of objects into the population, and $\bar\ell$ is the average lifetime of objects in the steady-state.  In this work we use the integrations described briefly above, and in detail in \citet{Jedicke2025}, to calculate the average flux and lifetime of objects in co-orbital motion with Earth.  The diameter distribution of the co-orbitals is calculated by accounting for the speed, flux, diameter, and density distribution of objects that impact the lunar surface, the diameter of the crater created by the impact, and the size and speed of the material ejected from the crater.  The ejecta are then integrated as described in \S\ref{ss.Method_co-orbital_dynamics}.

%##################################################
\section{Results \& Discussion}
\label{sec:results}

%---------------------------------------------
\subsection{Known co-orbitals}
\label{ss.R+D_known_co-orbitals}

We applied the procedure described above to the set of known\footnote{Those in NASA's JPL Small-Body Database at \url{https://ssd.jpl.nasa.gov/tools/sbdb_query.html} as of 2025 Nov 19.} small bodies with $a\in[0.97:1.03]\au$ and identified those that exhibit co-orbital behavior in the years 1600-2500 (tables~\ref{tab.Earth_co-orbitals_known_1600_2500_NOW} and \ref{tab.Earth_co-orbitals_known_1600_2500_NOT_NOW}, fig.~\ref{fig.e_vs_i_lunar}).
We identified almost 100 objects, most of them in the horseshoe regime or transitioning between co-orbital regimes involving a horseshoe, while \nKnownCoorbitalsNOW\ are currently co-orbital. 
The set of objects is not the same as those identified by \citet{DiRuzza2023} because we have used a different procedure, an updated set of candidates, and have included objects with inclinations $>10\arcdeg$.

The definition of co-orbital means that all the known objects have semi-major axes of $\sim1\au$ so it is not surprising that their average semi-major axis is $\fAvgSemiMajorAxisAU\au$. The objects have a mean eccentricity of $\bar{e}\sim\fAvgEccentricity$ and mean inclination of $\bar{i}\sim\fAvgInclination\arcdeg$ but there are a small number of objects with $e>0.4$ and $i>20\arcdeg$ which will be discussed below.  With a median absolute magnitude of $H=\fMedianAbsMag$, corresponding to a diameter\footref{ftn.albedo} of only $d=\fMedianDiameterMeters\meter$, these objects are difficult for contemporary asteroid surveys to detect and $\nKnownCoorbitalsTenM$ of them are $>10\meter$ diameter\footref{ftn.albedo} (\fig{fig.known_coorbital_absmag}).

\begin{center}
    \begin{minipage}{12cm}  % Adjust the width as needed
        \includegraphics[width=12cm]{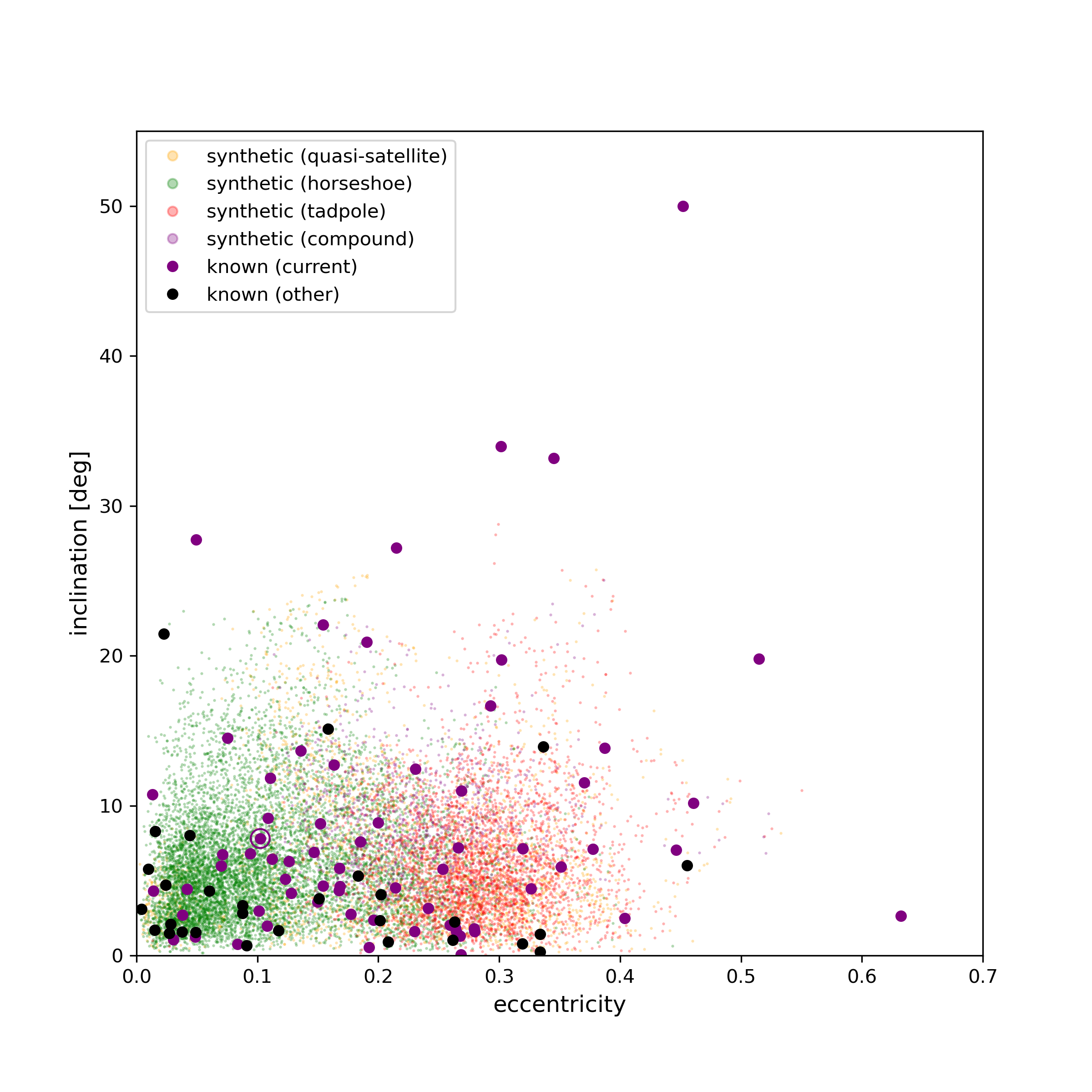}
        \captionof{figure}{
            The time-averaged eccentricity and inclination of synthetic Earth co-orbitals that were lunar ejecta over the duration of their time spent in each co-orbital regime.
            (purple circles) The present-day values of the orbital elements for known asteroids currently in co-orbital motion with Earth (\tab{tab.Earth_co-orbitals_known_1600_2500_NOW}) and
            (black circles) known asteroids that were, or will be, co-orbital with Earth in the years 1600-2500 (\tab{tab.Earth_co-orbitals_known_1600_2500_NOT_NOW}). 
            The point representing \Kamo\ is circled.}
        \label{fig.e_vs_i_lunar}
    \end{minipage}
\end{center}

\begin{center}
    \begin{minipage}{15cm}  % Adjust the width as needed
        \includegraphics[width=\columnwidth]{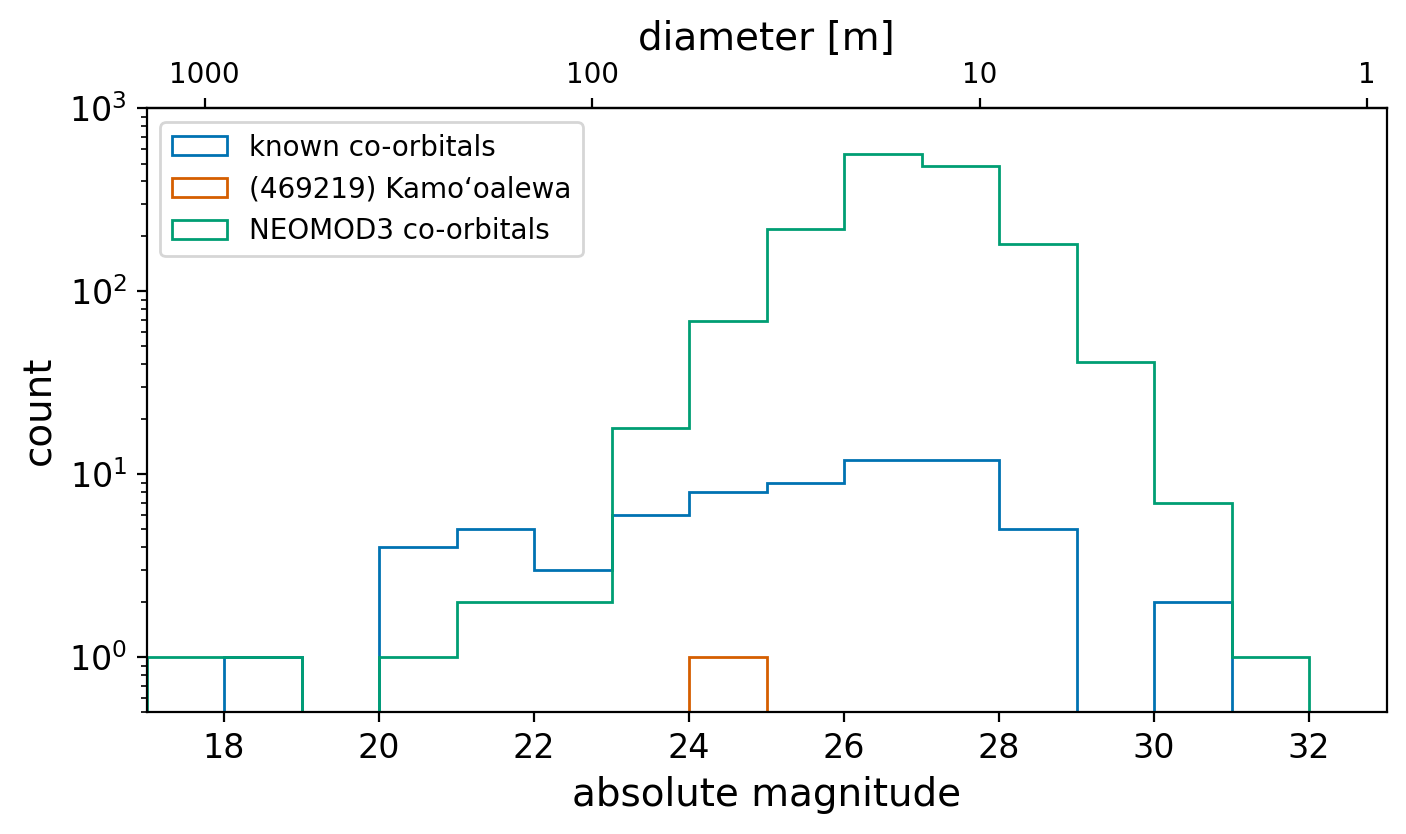}
        \captionof{figure}{
            The absolute magnitudes and diameters of Earth's known co-orbitals and the unbiased synthetic population of co-orbitals generated with NEOMOD3 assuming that the objects have albedos of $0.143$ such that $H=17.75$ corresponds to a diameter of $1\km$.}
        \label{fig.known_coorbital_absmag}
    \end{minipage}
\end{center}

\begin{table}[ht]
    \centering
    \newsavebox{\mytab}
    \sbox{\mytab}{%
        \begin{tabular}{rrrrrr}
            \toprule
                             &                  &                          &                           & nominal                         &   \\
                             &                  &                          &                           & lunar                           &   \\
                             & known            & known                    & known                     &  model                          & NEOMOD3   \\
                      regime &                  & $>10\meter$              & $H<26$                    & $>10\meter$                     & $>10\meter$ \\
            \midrule
            % the values in the following 4 lines are created with
            % import Earth_co-orbitals as coorb
            % coorb.table_known_coorbital_regimes()
                         any & \nKnownCoorbNOWANY & \nKnownCoorbNOWANYTenMeter & \nKnownCoorbNOWANYHTwoSix & \nCumulativeCoorbitalsTenMeterANY & \nNEOMODcoorb$\pm40$ \\
             quasi-satellite & \nKnownCoorbNOWQS & \nKnownCoorbNOWQSTenMeter & \nKnownCoorbNOWQSHTwoSix & \nCumulativeCoorbitalsTenMeterQS  & $318\pm18$ \\
                   horseshoe & \nKnownCoorbNOWHS & \nKnownCoorbNOWHSTenMeter & \nKnownCoorbNOWHSHTwoSix & \nCumulativeCoorbitalsTenMeterHS  & $380\pm19$  \\
                     tadpole & \nKnownCoorbNOWTP & \nKnownCoorbNOWTPTenMeter & \nKnownCoorbNOWTPHTwoSix & \nCumulativeCoorbitalsTenMeterTP  &  $783\pm28$ \\
                    compound & \nKnownCoorbNOWCP & \nKnownCoorbNOWCPTenMeter & \nKnownCoorbNOWCPHTwoSix & \nCumulativeCoorbitalsTenMeterCP  & $108\pm10$  \\
                    \bottomrule
        \end{tabular}%
    }
    \captionsetup{width=\wd\mytab}
    \caption{The number of known current co-orbitals, those of $\>10\meter$ diameter\protect\footref{ftn.albedo}, and those with absolute magnitude $H<26$ currently in the specified regime, our nominal prediction for the number of co-orbitals in each regime with a lunar origin, and the total number of co-orbitals expected from the main belt using NEOMOD3 \citep{Nesvorny2024-NEOMOD3}.  The number of objects in the lunar model in any regime is not the sum of the number in the four separate regimes because the regimes have different average durations (\ie, \eqn{eqn.Nany}).  Uncertainties are not included in the nominal lunar model because they are dominated by systematics as discussed in \S\ref{ss.SystematicUncertainties}. The uncertainties on the NEOMOD3 model are simply $\sqrt{N}$.}
    \label{tab.Ncoorbital_per_regime}
    \usebox{\mytab}
\end{table}

We identified 3 objects in tadpole orbits, \designationsub{2005}{UH}{6} ($L_5$ Trojan), \designationsub{2005}{QQ}{87} ($L_5$ Trojan), and \designationsub{2024}{JR}{16} ($L_4$ Trojan), in addition to the 2 widely known objects, \designationsub{2010}{TK}{7} and \designationsub{2020}{XL}{5} (both $L_4$ Trojans), and verified the synodic behavior of the three objects with the NEOCC visualization tool\footnote{\url{https://neo.ssa.esa.int/orbit-visualiser}}.  \designationsub{2005}{UH}{6} has a perihelion distance of $0.37\au$ and an aphelion distance of $1.63\au$ so it crosses the orbits of Venus, Earth and Mars, which causes the asteroid's resonant angle to oscillate around a variable value. With an eccentricity of $0.35$, \designationsub{2024}{JR}{16} suffers from similar dynamical effects. These two objects were dubbed `jumping' by \citet{Pan2025} and  both have orbits with resonant angles that oscillate around a value higher than $100\arcdeg$ in absolute value due to their high eccentricities \citep{2017PoRoVi}. In addition, \designationsub{2005}{QQ}{87}'s orbital inclination of $34\arcdeg$ contributes to its unusual dynamical evolution (\cf\ fig.~6f in \cite{Christou00}).

\begin{table*}[t]
\centering
\caption{NEOMOD3-derived main belt source region probabilities for known Earth co-orbitals and the average source region probabilities for the \nKnownCoorbitals\ known Earth co-orbitals identified in this work.  NEOMOD3 source regions with no values $>0.015$ have been omitted.  The source regions are the $\nu_6$ secular resonance, the 3:1, 5:2, and 8:3 mean motion resonances, the inner belt region, and the Hungaria and Phocaea regions.}
\label{tab.known_coorbital_source_probabilities}
\begin{tabular}{R{5cm}cccccccc}
\toprule
object & $p_{\mathrm{nu6}}$ & $p_{31}$ & $p_{52}$ & $p_{83}$ & $p_{115}$ & $p_{\mathrm{inner}}$ & $p_{\mathrm{Hun}}$ & $p_{\mathrm{Pho}}$ \\
\midrule
3753 Cruithne (1986 TO) & 0.11 & 0.32 & 0.03 & 0.14 & 0.06 & 0.27 & 0.04 & 0.01 \\
85770 (1998 UP1)        & 0.42 & 0.39 & 0.01 & 0.02 & 0.02 & 0.08 & 0.05 & 0.01 \\
255071 (2005 UH6)       & 0.35 & 0.26 & 0.06 & 0.02 & 0.00 & 0.28 & 0.04 & 0.00 \\
706765 (2010 TK7)       & 0.53 & 0.20 & 0.01 & 0.02 & 0.00 & 0.21 & 0.03 & 0.00 \\
(2013 LX28)             & 0.70 & 0.20 & 0.00 & 0.01 & 0.00 & 0.04 & 0.03 & 0.02 \\
average co-orbital      & 0.78 & 0.16 & 0.00 & 0.00 & 0.00 & 0.05 & 0.01 & 0.00 \\\bottomrule
\end{tabular}
\end{table*}

Assuming that the co-orbitals derive from the main belt per NEOMOD3 \citep{Nesvorny2024-NEOMOD3} their orbital elements suggest that they have a 78\% probability of being delivered to near-Earth space by the $\nu_6$ resonance, a 16\% chance of being processed through the 3:1 resonance, or a 5\% likelihood of a provenance in the forest of minor resonances in the inner region of the belt (\tab{tab.known_coorbital_source_probabilities}).  The other nine sources considered by \citet{Nesvorny2024-NEOMOD3} contribute $\lesssim 1$\% each.  Given that the 3:1 resonance marks the outer boundary of the inner belt and, for the sake of argument, assuming that half of the objects delivered from that resonance to near-Earth space were originally interior to the resonanace, then about 97\% of Earth's known co-orbitals derive from the inner belt.  If this is the case then most of these objects should be S-complex objects \citep[\eg][]{DeMeo2014}.

Of the \nKnownCoorbitals\ known Earth co-orbitals identified in this work only 5 have \emph{any} NEOMOD3 source probability $>1.5$\% for any source region other than the $\nu_6$ and 3:1 resonances and the `inner' sources (\tab{tab.known_coorbital_source_probabilities}).  Even so, the inner belt remains the most likely source of these 5 objects with the highest non-inner belt source region probability of 14\% for 3753 Cruithne (1986 TO) originating in the 8:3 resonance at about $2.71\au$.  In that region of the belt S- and C-complex asteroids are found in roughly equal proportion \citep{DeMeo2014}.

%vconfirmed also from NEOCC orbit visualization, but very displaced wrt the usual +/- 60deg configuration. Pan and Gallardo do not classify it probably because it has a large eccentricity and the tadpole behavior is not regular, because it is likely affected by a 4th body (remember the comment above on the high eccentric objects).

%2005 QQ87: very inclined, so TP spans a wide range of resonant angle (from -20 to -160 approx), but the phase space is ok and also from NEOCC. For Pan and Gallardo this is classified as "jumping", that is its center of libration is moving. This is due to the fact that the orbit is very inclined.

%2024 JR16: confirmed by synodic visualization from NEOCC, large range of resonant angle (50,150), not so inclined, but quite eccentric. Also this one is classified as "jumping" in Pan and Gallardo. In this case, the issue can be the highly eccentric orbit.

%Discuss \tab{tab.Ncoorbital_per_regime}.
%why did we find \nKnownCoorbNOWTPHTwoSix\ TP, while officially there are 2? In the approach I used for the whole paper for me tadpole are any orbits that can stem from L4/L5 (as stated in the introduction) and not all the orbits of  the corresponding family have a libration center at 60 degree. It depends on the eccentricity of the orbits, as shown in Pousse et al. 2017. %What is the citation for the official number 2? HERE: \cite{Hui} and https://www.nature.com/articles/s41467-022-27988-4 I HAVE PUT THIS IN THE INTRODUCTION

%---------------------------------------------
\subsection{Synthetic main belt co-orbitals}
\label{ss.R+D_synthetic_mainbelt_co-orbitals}

This work was motivated by recent spectroscopic studies of known co-orbitals that suggest they originated on the lunar surface, so it is also important to investigate the number of co-orbitals that can be explained through their dynamical evolution from the main belt.  To that end we employed NEOMOD3 \citep{Nesvorny2024-NEOMOD3} to generate a pseudo-realistic population of objects that are $\ge10\meter$ diameter with semi-major axes in the range $0.97\au \le a \le 1.03\au$.  All the objects generated by NEOMOD3 have a main belt or Jupiter family comet (JFC) provenance and the model provided \nNEOMODgenerated\ objects with no restrictions on eccentricity and inclination.  NEOMOD3 does not generate the other three angular orbital elements, the longitude of the ascending node, the argument of perihelion, and the mean anomaly, so we generated each of them in a uniform random distribution in the range $[0\arcdeg,360\arcdeg)$.

Using REBOUND we integrated those objects forward and backward by 75,000$\Days$, about $205\yr$, in each direction and then identified \nNEOMODcoorb\ of the objects as being co-orbital at the current epoch compared to the \nKnownCoorbitalsNOWTenM\ known co-orbital objects larger than $10\meter$ diameter\footref{ftn.albedo} (\tab{tab.Ncoorbital_per_regime}).  Taken at face value, and assuming that all the known co-orbitals originated in the main belt or JFC population, it implies that $\lesssim 4$\% of the co-orbitals $>10\meter$ diameter have been discovered.  On the other hand, NEOMOD3 predicts that there are \nNEOMODcoorbHundredM\ co-orbital objects $>100\meter$ diameter while there are already \nKnownCoorbitalsNOWHundredM\ known in this size range.  We speculate that the excess may be due to errors in NEOMOD3, perhaps due to it not including tidal disruptions of asteroids that pass close to Earth \citep[\eg][]{Granvik2024-NEOTidalDisruption}, neglecting a population of objects that originate on the lunar surface, and/or misestimating the observational bias in detecting the smallest asteroids.

The number of co-orbital objects predicted by NEOMOD3 that are similar to \Kamo\ is of particular interest given that there are claims that it has a lunar-like spectrum \citep{Sharkey2021-2016HO3,Zhu2025} and \citet{Jiao2024-Kamooalewa} specifically link it to the lunar crater Giordano Bruno.  We define \Kamo-like objects as co-orbitals with eccentricities within $\pm0.05$ and inclinations within $5\arcdeg$ of \Kamo's orbital elements for which there are \nCoorbitalsLikeKamo\ objects in our NEOMOD3 instantiation.  Constraining the similarity specifically to quasi-satellites with $H<27$, \Kamo's $H=24.3$, suggests that the main belt and JFC sources can generate about \nQuasiSatellitesLikeKamo\ \Kamo-like co-orbitals so that invoking a lunar source for this co-orbital is not necessary on dynamical grounds. This is in agreement with \citet{Fenucci2026} who specifically considered the formation of quasi-satellites with a main belt provenance and found that it could produce $1.23\pm0.13$ \Kamo-like objects.

\begin{table}[ht]
    \centering
    \newsavebox{\coorbtab}
    \newsavebox{\knownCoorbTable}
    \sbox{\knownCoorbTable}{%
    \begin{tabular}{rrlcrcl}
        \toprule
        number  & name      & designation                       & H & diameter & taxonomic    & references  \\
                &           &                                   &   &   [m]    & class\tss{a} & [\eg] \\
        \midrule
        % the following lines are produced with coorb.extract_likely_mainbelt_co-orbitals()
		   3753  &   Cruithne &    \designation{1986}{TO}		 & 15.4 & 2924 & Q             & \cite{Wiegert1997Cruithne} \\
		   85770 &            & \designationsub{1998}{UP}{1}	 & 20.7 &  255 & S             & \cite{Galiazzo2014} \\
		255071  &            & \designationsub{2005}{UH}{6}	    & 18.4 &  755 & n/a           & \cite{Birlan2010-160NEOS-EURONEAR} \\
		        &            & \designationsub{2005}{QQ}{87}	& 22.8 &   96 & n/a           & \cite{Tardioli2017-NEO-obliquity} \\
		        &            & \designationsub{2013}{LX}{28}	& 21.9 &  149 & n/a           & \cite{Connors2014-2013LX28-quasi-satellite} \\
		        &            & \designationsub{2016}{CA}{138}	& 23.3 &   77 & n/a           & \cite{Borisov2023} \\
		        &            & \designationsub{2017}{XQ}{60}	& 24.5 &   45 & n/a           & \cite{Kaplan2020-HorseshoeCoorbitals} \\
        \bottomrule
    \end{tabular}
    }
    \captionsetup{width=\wd\knownCoorbTable}
    \caption{
        Selected properties of known Earth co-orbitals with $e>0.5$ and $i>25\arcdeg$ which are likely to have a main belt provenance in order of discovery year.  The absolute magnitude, $H$, is from JPL Horizons\tss{b} and the diameter is calculated from $H$ assuming that all objects have an albedo of $0.143$ which maps $H=17.75$ to a diameter of $1000\meter$.  Taxonomic classes and citations are from ESA's near-Earth objects coordination centre.\tss{c} We provide one reference to other works that mention these objects.}
    \label{tab.candidate-non-lunar-co-orbitals}
    \usebox{\knownCoorbTable}
    \begin{minipage}{\wd\knownCoorbTable}
        \usebox{\coorbtab}
        \par\vspace{2pt}
        \footnotesize
        \textsuperscript{a}\citet{DeMeo2014-MarsEncounters},
        \textsuperscript{b}\url{https://ssd.jpl.nasa.gov/horizons},\\
        \textsuperscript{c}\url{https://neo.ssa.esa.int/search-for-asteroids}\\
    \end{minipage}
\end{table}

The eccentricity and inclination distributions of the Earth co-orbitals generated with NEOMOD3 cover the entire range of values of the known co-orbitals (\fig{fig.e_vs_i_neomod3}) suggesting that the main belt and JFC are viable source regions for these objects.  The average eccentricity and inclination of the NEOMOD3 co-orbitals are $\fNEOMODavgE$ and $\fNEOMODavgI\arcdeg$ respectively.  About 50\% of the NEOMOD3 co-orbitals are on tadpole orbits in contrast with the $\sim$14\% of known objects with $H<26$. These tadpoles have oscillation periods in their resonant angles of about 400 years and the resonant angle is shifted towards $\pm100\arcdeg$, like \designationsub{2005}{UH}{6}, \designationsub{2005}{QQ}{87}, and \designationsub{2024}{JR}{16} (\tab{tab.candidate-non-lunar-co-orbitals} and \tab{tab.Earth_co-orbitals_known_1600_2500_NOT_NOW}). This is likely an observational selection effect since the minimum resonant angle for a tadpole is $25\arcdeg$ so the minimum distance between Earth and an object is $\sim0.43\au$ when it is only $77.5\arcdeg$ from the Sun, an area that is difficult for asteroid surveys to image.  Furthermore, an $H=26$ object has an apparent magnitude of $V\sim26.9$, about 5~magnitudes fainter than the \PS\ limiting magnitude, the most sensitive dedicated, long-term asteroid survey to-date \citep{Denneau2013}.  An object would need to have an absolute magnitude of $H\lesssim21$ to be brighter than the \PS\ limiting magnitude even when surveying in the bright sky in the direction of morning or evening twilight.  Co-orbital objects in the other 3 regimes can experience periods when their resonant angle is near zero which brings them closest to Earth during which time they are more likely to be discovered. Thus, we predict that there is a large population of undetected tadpole co-orbitals and that the Vera Rubin Observatory will detect many during its twilight survey campaign \citep[\eg][]{Schwamb2023-TuningLSSTforSSS,Bolin2025}, in agreement with the prospects for discovery of lunar ejecta calculated by \citet{Wu2026}.

\begin{center}
    \begin{minipage}{10cm}  % Adjust the width as needed
        \includegraphics[width=\columnwidth]{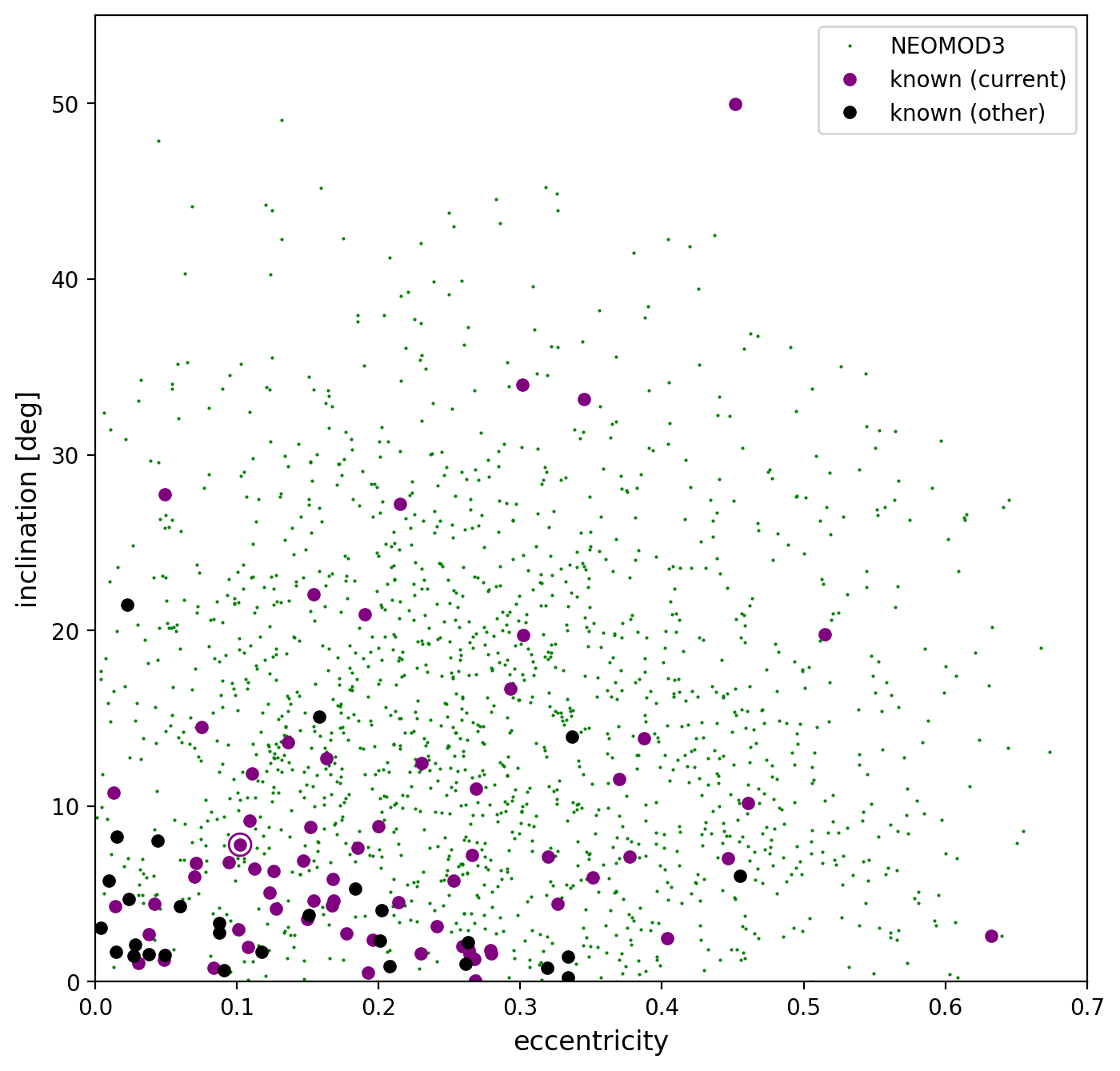}
        \captionof{figure}{
            The eccentricity and inclination of synthetic Earth co-orbitals that have a main belt or JFC provenance according to NEOMOD3.
            (purple circles) The present-day values of the orbital elements for known asteroids currently in co-orbital motion with Earth (\tab{tab.Earth_co-orbitals_known_1600_2500_NOW}) and
            (black circles) known asteroids that were, or will be, co-orbital with Earth in the years 1600-2500 (\tab{tab.Earth_co-orbitals_known_1600_2500_NOT_NOW}). 
            The point representing \Kamo\ is circled.}
        \label{fig.e_vs_i_neomod3}
    \end{minipage}
\end{center}

%\todo{Rob}{Just based on dynamics is there an expectation for which would be the most populated regime? For this sentence that you wrote in the following section "In the available phase space in the $(\theta,e)$ plane we would expect that HS co-orbitals would be the least populated with an eccentricity $\gtrsim0.5$, the QS would be the next most likely regime with $e>0.5$, and the tadpoles would be the most populous with a mean $e<0.5$.". In particular, you computed that the average eccentricity of the main belt co-orbital is 0.28.
%Based on dynamics and observational bias is there an explanation for why about half of the known objects with $H<26$ are compound? The same that I wrote in another answer/comment: they behave almost like HS, that is, they approach the Earth more than others and are easier to detect (at least you told me that when I was preparing the Di Ruzza paper).
%We should not discuss the lunar and NEOMOD3 models here but there are big differences between the known $H<26$ population and the two models that we need to discuss/anticipate below.}

%---------------------------------------------
\subsection{Synthetic lunar co-orbitals}
\label{ss.R+D_synthetic_lunar_co-orbitals}

We then applied the procedure for identifying periods of co-orbital motion to our synthetic population of lunar ejecta while in heliocentric orbit.

The eccentricity and inclination distributions of the lunar ejecta co-orbitals (\fig{fig.e_vs_i_lunar}) are different from the Earth co-orbitals in NEOMOD3 (\fig{fig.e_vs_i_neomod3}) --- they are more concentrated at low-$e$ and low-$i$ than the main belt/JFC co-orbital population.  The maximum eccentricity is $\synLunarMaxe$ and the maximum inclination is $\synLunarMaxi\arcdeg$, \ie\ all the lunar ejecta co-orbitals are on heliocentric prograde orbits.  This could be an artifact of the method described in \citet{Jedicke2025} that removes objects that `leave the `extended intermediate source region' (xISR) defined by heliocentric semi-major axes in the range $a\in[0.8\au,1.2\au]$, eccentricities in the range $e\in[0,0.2]$, and inclinations in the range $i\in[0\arcdeg,5\arcdeg]$' but we think it is unlikely that an object's orbital elements can evolve outside the xISR and then evolve onto a retrograde orbit with $a\sim1$.  While retrograde orbits exist in the circular restricted three-body problem \citep{Morais2013-retrograde-coorbitals} and have been identified in association with Jupiter and Saturn \citep{Morais2013-Jupiter+Saturn-retrograde-coorbitals, Wiegert2017} none exist amongst Earth's known co-orbitals.  

\begin{figure}
    \centering
    \includegraphics[width=0.49\columnwidth]{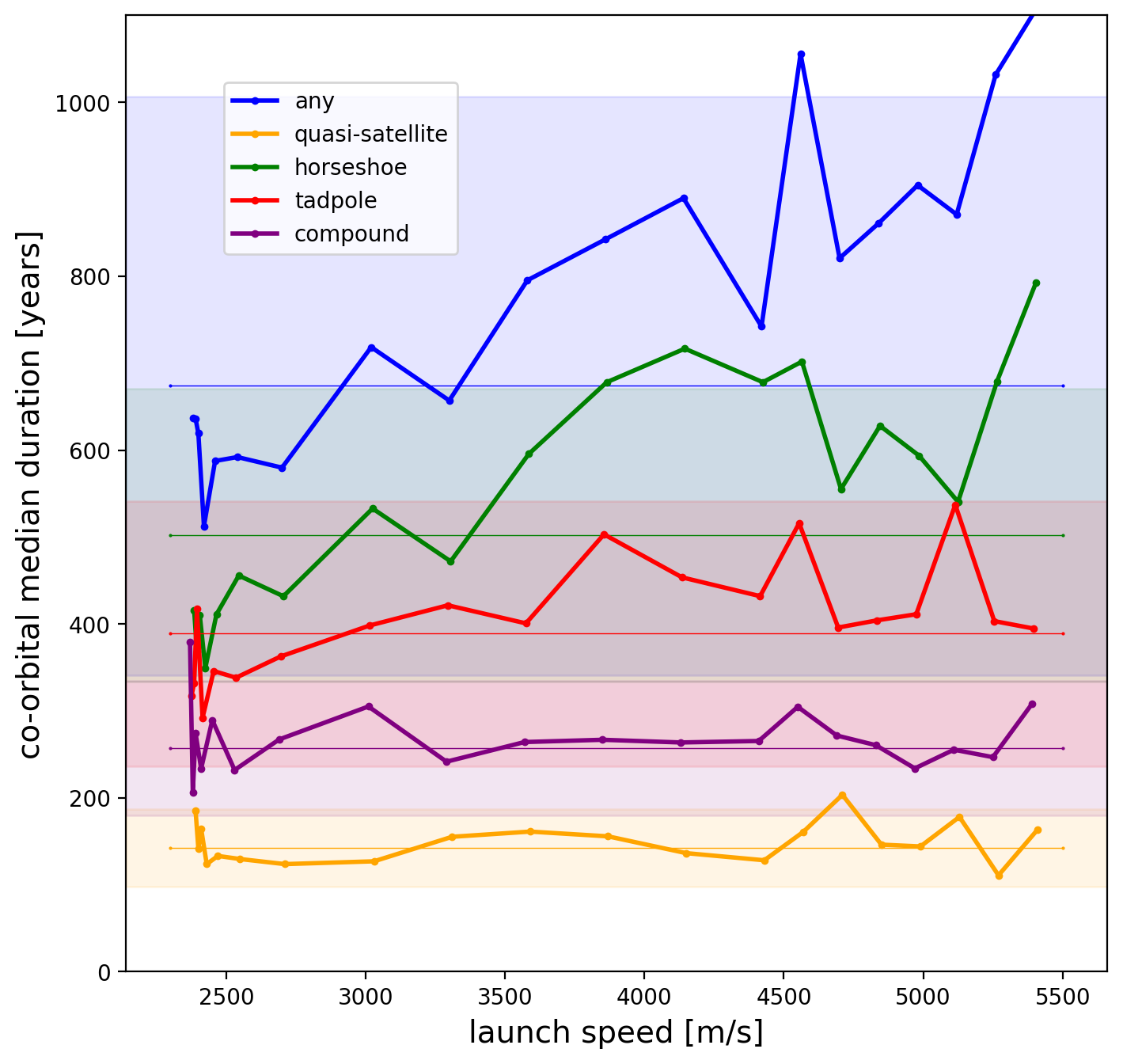}
    \includegraphics[width=0.49\columnwidth]{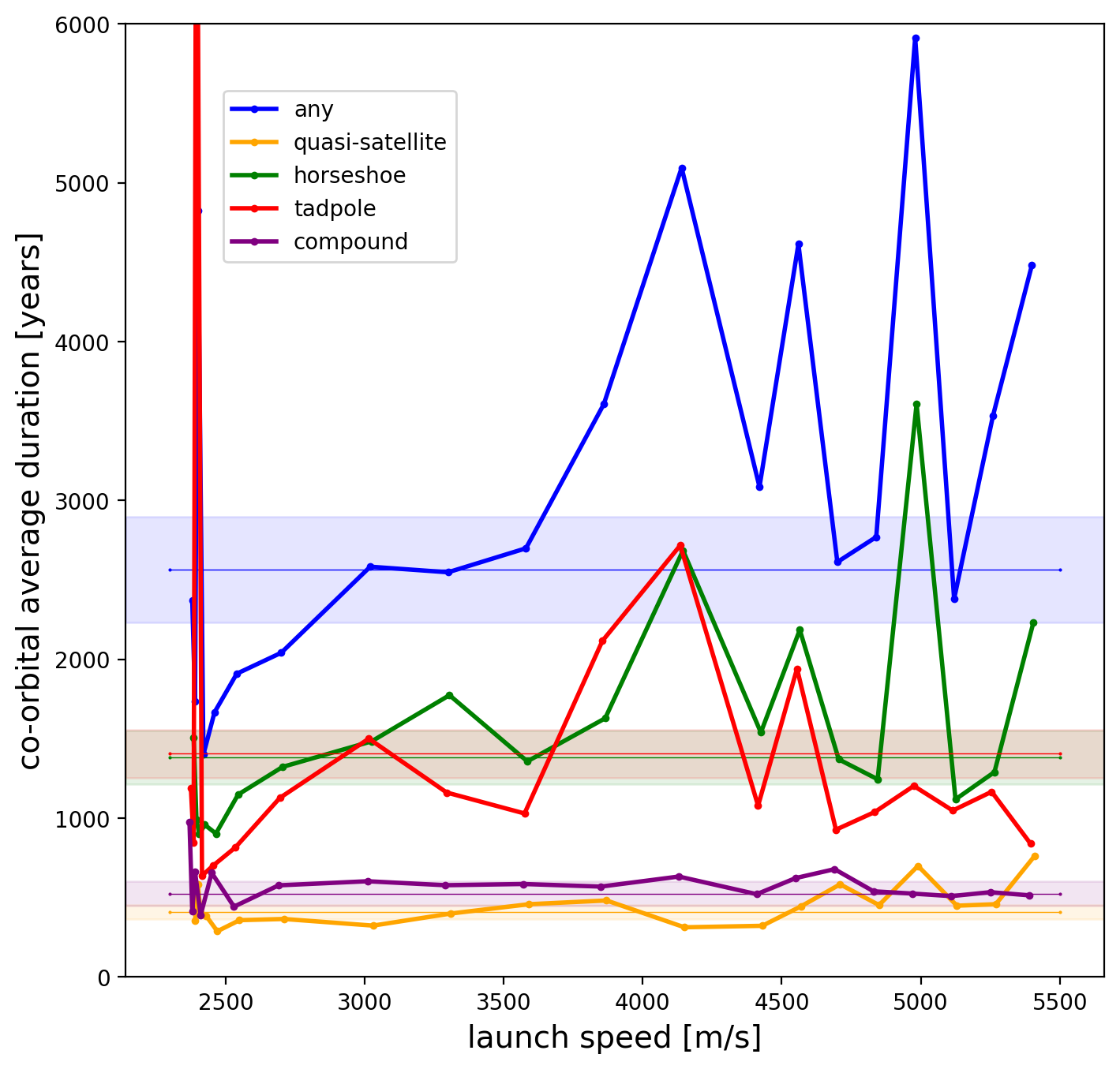}
    \caption{
        ({\bf left}) The points connected by solid piecewise lines represent the median duration of co-orbital motion in each of the four regimes and any regime as a function of launch speed from the lunar surface.  The data points are offset by $(0,\pm5,\pm10)\mps$ from their actual values to improve readability.  The uncertainty on the duration at each launch speed for each regime is not shown for clarity. Thin solid horizontal lines represent the weighted mean duration of the median values for the regime while the shaded bands represent the standard error on the means.
        ({\bf right}) Similar to the left panel except for the average duration of co-orbital motion instead of the median.  Note the different scales on the $y$-axis.}
\label{fig.speed_duration}
\end{figure}

\begin{table}[ht]
    \centering
%    \newsavebox{\coorbtab}
    \sbox{\coorbtab}{%
    \begin{tabular}{
      l
      S[table-format=1.3(3)]
      c
      S[table-format=2.2(2)]
      r
    }
        \toprule
              & \multicolumn{2}{c}{eccentricity}                              & \multicolumn{2}{c}{inclination [deg]} \\
        class & \multicolumn{1}{c}{mean$\pm\sigma$} & \multicolumn{1}{c}{RMS} & \multicolumn{1}{c}{mean$\pm\sigma$} & \multicolumn{1}{c}{RMS} \\
        \midrule
        % The following 4 lines are produced by Earth_co-orbitals.plot_lunar_ejecta_i_e()
        quasi-satellite  &  0.215 \pm 0.002  &  \num{0.24} &  6.94 \pm 0.10  &  \num{ 8.6} \\
              horseshoe  &  0.111 \pm 0.001  &  \num{0.13} &  6.24 \pm 0.04  &  \num{ 7.4} \\
                tadpole  &  0.284 \pm 0.001  &  \num{0.29} &  6.41 \pm 0.07  &  \num{ 7.5} \\
               compound  &  0.228 \pm 0.002  &  \num{0.24} &  9.28 \pm 0.12  &  \num{10.0} \\
        \bottomrule
    \end{tabular}
    }
    \captionsetup{width=\wd\coorbtab}
    \caption{The average eccentricity and inclination and RMS of the distributions for synthetic lunar ejecta for each of the Earth co-orbital classes. The uncertainties are the standard error on the mean. The values correspond to all the particles in our integrations that became co-orbital; they are not weighted by the number of co-orbitals generated at each speed.}
    \label{tab.synthetic-coorbital-e+i}
    \usebox{\coorbtab}
\end{table}

In the quasi-coplanar approximation each co-orbital regime is associated with a different range in eccentricity and resonant angle \citep[\eg][]{Pousse+Alessi2022, Cortese2025} and, assuming that the $(e,\theta)$ phase-space is uniformly populated, we would expect that the tadpoles would be the most populous with a mean $e<0.5$, the QS would be the next most likely regime with $e>0.5$, and the HS co-orbitals would be the least populated.  
These expectations are modified by the initial conditions of the synthetic lunar ejecta which preferentially generate co-orbitals with low eccentricity.  Furthermore, particles with $e>0.5$ are likely to have close encounters with Venus and Mars so they will be short-lived and not contribute much to the population statistics.  Thus, our results are differ from our naive expectations.
There are statistically significant differences between the mean eccentricities of the synthetic lunar co-orbitals in the different regimes with the horseshoe co-orbitals having the lowest $e$ while the tadpoles have a value about $2.5\times$ larger (\tab{tab.synthetic-coorbital-e+i}). 
There are also statistically significant differences in the mean inclinations of the co-orbital particles but the compound regime's inclination is $\gtrsim15\sigma$ from the other three regimes because they can only exist in the inclined case.

The \emph{biased} $(e,i)$ distribution of known objects that undergo co-orbital episodes between years 1600 and 2500 (\tab{tab.Earth_co-orbitals_known_1600_2500_NOW} and \tab{tab.Earth_co-orbitals_known_1600_2500_NOT_NOW}) closely matches the corresponding distribution for the synthetic lunar ejecta. This agreement suggests that a lunar origin for at least some known Earth co-orbitals is plausible (\fig{fig.e_vs_i_lunar}). Seven objects are notable outliers from this match and are more consistent with a main-belt origin (\tab{tab.candidate-non-lunar-co-orbitals}).  It is no surprise that the largest of the seven, the first identified Earth co-orbital, \namedasteroid{3753}{Cruithne} \citep{Wiegert1997Cruithne}, was discovered 12 years earlier than the second object, and that the absolute magnitudes of the objects increase with discovery year because they are more difficult to detect and required the advent of superior asteroid surveys.  Its size alone would suggest that it is unlikely to be lunar ejecta and it is consistent with being the largest of the co-orbitals with a main belt provenance (\tab{fig.known_coorbital_absmag}).  Only the first two discovered objects have taxonomic classifications and both are consistent with a main belt origin. The Q-class \namedasteroid{3753}{Cruithne} is indicative of a fresh surface on an ordinary chondrite-like body while the S-class \designationsub{1998}{UP}{1} is the more weathered version \citep[\eg][]{Jedicke2004,Binzel2010} of the same type of object, so that both objects can be associated with a main belt origin.  Furthermore, it has been suggested that \designationsub{1998}{UP}{1} is likely to have been a Hungaria asteroid, a population of objects found on the inner edge of the main belt \citep{Galiazzo2014}.  While the other objects have not been assigned taxonomies the `[r]ecovery prospects for [\designationsub{2013}{LX}{28}] are excellent' \citep{Connors2014-2013LX28-quasi-satellite}, so we expect that this object and others will be utilized to test the provenance of Earth's co-orbitals.

We note that four of the co-orbitals in \tab{tab.candidate-non-lunar-co-orbitals}, objects that are more likely of main belt origin according to our analysis, also appear in the table of five co-orbitals with possible origins beyond the inner main belt according to NEOMOD3 (\tab{tab.known_coorbital_source_probabilities}).  The co-orbital \designationsub{1998}{UP}{1} is the most likely of the five to originate in the Hungaria population with a 5\% probability which provides some support for \citet{Galiazzo2014}'s hypothesis.

The duration of co-orbital motion of the particles in our simulation ranged from tens of years to a $\Myr$ but the most likely durations were in the hundreds to few thousand year range.  The median durations in each regime are well behaved as a function of launch speed and are either flat or increase with launch speed (\fig{fig.speed_duration}\ left).   We provide the median durations to illustrate that the behavior of the durations are more regular than suggested by the average durations which are necessary for calculating the steady-state population (\eqn{eqn.nSteadyState-basic} and \S\ref{ss.R+D_co-orbital_steady_state_SFD}).  The average durations of co-orbital motion are about a few times larger than the median durations but with high RMS due to the statistics of a small number of long duration events (\fig{fig.speed_duration}\ right).  Note that the average duration of `any' co-orbital, $\mu_{any}$, is not the sum of the averages of the individual regimes but is given by
\begin{equation}
    \mu_{any} = \frac{1}{n_{any}}
    \left(n_{QS}\mu_{QS}+n_{HS}\mu_{HS}+n_{TP}\mu_{TP}+n_{CP}\mu_{CP}\right),
    \label{eqn.Nany}
\end{equation}
where $n_{any}$ is the number of continuous co-orbital intervals, \ie\ such that the transitions from one  regime to another are not interrupted by non-resonant intervals, $n_X$ is the number of intervals corresponding to regime $X$, and $\mu_X$ denotes the average duration in the regime.

\begin{figure}[ht]
    \includegraphics[width=0.465\columnwidth]{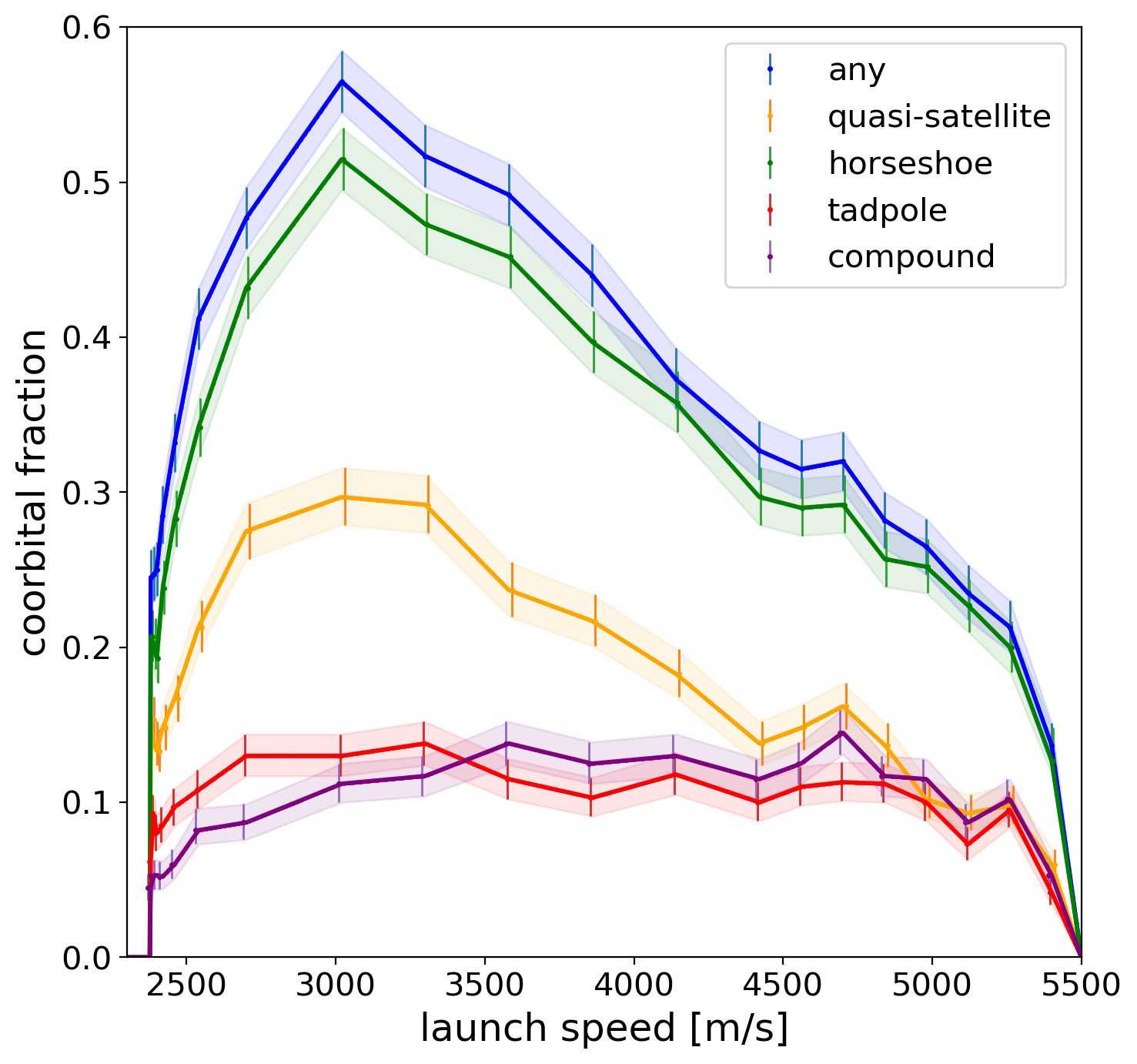}
    \includegraphics[width=0.49\columnwidth]{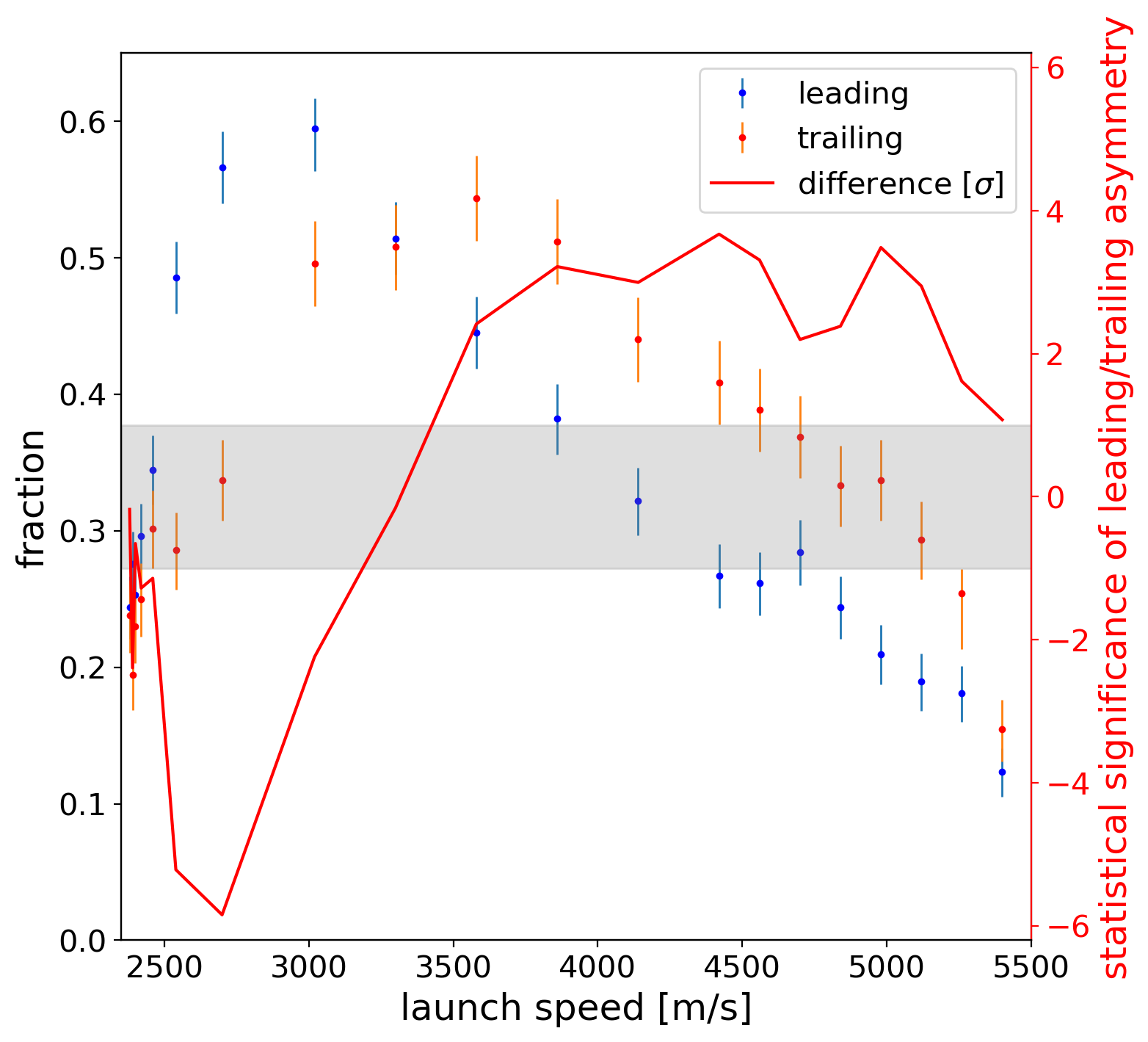}
    \caption{
        ({\bf left}) The average fraction of particles that experience different, or any, regimes of co-orbital motion as a function of their lunar ejection speed. The solid lines are a piecewise-continuous linear function that connect the data points.  The coloured bands are the $\pm1$-sigma regions that are used in our study of the systematic uncertainty on the co-orbitals's steady-state size-frequency distribution.  The functions representing the average fractions are forced to zero for launch speeds $<2380\mps$ and for speeds $\ge5500\mps$. 
        ({\bf right}) The fraction of particles that experience at least one co-orbital period as a function their launch speed for particles ejected from leading ($<0$) or trailing lunar longitudes ($\ge0$).  The right hand $y$-axis provides the statistical significance of the difference between the leading/trailing asymmetry.}
    \label{fig.coorbital_fractions}
\end{figure}

We fit the average durations as a function of launch speed to straight lines and found that two of the four regimes are consistent with being constant and the other two were within $3\sigma$ of being flat. Particles in `any' regime also exhibit $<3\sigma$ deviation from a linear trend with launch speed.  Given that the median durations are well behaved, the average durations are consistent with being constant within $3\sigma$, and we will show below that the systematic uncertainties in the lunar ejecta size-frequency distribution overwhelm our assumptions on the relationship between co-orbital duration and launch speed.  We used the weighted mean over all launch speeds within a regime as the canonical duration time for objects regardless of their launch speed.  We will employ the standard error on the mean for our systematic study of the uncertainties on the steady-state size-frequency distribution in \S\ref{ss.SystematicUncertainties}. 

The thousand to a few 1,000~years average durations for our lunar co-orbitals is dramatically shorter than the 25,000~year lifetime of co-orbitals of main belt origin calculated by \citet{MoraisMorbidelli2002}.  While they predicted that there are $0.65\pm0.12$ $H<18$ Earth co-orbitals and $16.3\pm3.0$ with $H<22$, in good agreement with the 1 and 11 currently known respectively, we note that their source population is different from ours, aligned with the results in \S\ref{ss.R+D_synthetic_lunar_co-orbitals} and, more importantly, their output time was 100~years and they rejected cases shorter than 1,000~years.  This would have the effect of lengthening the average duration of their co-orbitals. They also noted that ``the typical eccentricities are quite high, that most objects have orbits which are Venus crossing ($e > 0.28$), and that many have orbits which are Mars crossing ($e>0.52$). ... typical inclinations can also be quite high and that most of the objects have orbits with $10\arcdeg < i < 45\arcdeg$'', in line with our analysis of co-orbitals with a main belt provenance in \S\ref{ss.R+D_synthetic_mainbelt_co-orbitals}.

We attributed that fact that the durations are independent of launch speed to the particles `forgetting' their launch conditions in the time between ejection and the co-orbital period. We then investigated whether the co-orbitals had a `memory' of their ejection speed revealed in their eccentricity distributions.  We found that the quasi-satellite, tadpole, and compound co-orbitals do not show significant correlations between launch speed, $v$, and their eccentricity during their co-orbital periods but the horseshoe co-orbitals do, with $\bar{e} = (0.0114\pm0.0008) v + (0.065\pm0.003)$.  We found an even stronger correlation between the mean eccentricity of $e<0.1$ QS co-orbitals\footnote{There is a gap in the quasi-satellites' eccentricity distribution at $e\sim0.1$ that we leave for examination to future work.} and their launch speeds of $\bar{e} = (0.017\pm0.001) v+(-0.018\pm0.003)$.  While the range of eccentricities is large at any launch speed these results suggest that high-$e$ HS co-orbitals are more likely to have been ejected at high speed and it would therefore be unusual to discover a large, high-$e$ HS co-orbital.

All the launch speeds we considered can generate co-orbitals and the fraction of particles that become co-orbital exceeds 50\% for launch speeds $\gtrsim2800\mps$ and $\lesssim3500\mps$ (\fig{fig.coorbital_fractions}).  The maximum fraction of $\sim57$\% occurs at a launch speed of $3020\mps$, the same value which maximizes the fraction of ejecta that become `temporarily bound objects', with a negative total energy with respect to the geocenter \citep{Jedicke2025}. \cite{Sfair2025} identified a peak in co-orbital production associated with an ejection speed of $2880\mps$.  While the horseshoe and quasi-satellite co-orbital regimes follow the same trend as objects in any co-orbital regime, the compound and tadpole co-orbitals display a different behaviour with a broad maximum fraction extending from about $2750\mps$ to over $5000\mps$.  \citet{CastroCisneros2023} found that only 6.6\% of their ejecta achieved co-orbital motion and those most likely to achieve co-orbital status were launched at `slightly above lunar escape' speed where our work suggests that the launch speed most likely to result in co-orbital motion is about $600\mps$ greater than the escape speed.  

We found that there is a launch speed dependent asymmetry in the fraction of test particles that become co-orbital from the Moon's leading and trailing hemispheres (\fig{fig.coorbital_fractions}, right) in contrast with \citet{CastroCisneros2023} who found that the trailing side generated more co-orbitals.  The actual asymmetry in the overall fraction of lunar ejecta that become co-orbital is then also a function of the number of particles ejected as a function of launch speed.  The statistical significance of the asymmetry at each launch speed is $>1\sigma$ at most of the speeds we examined and $>3\sigma$ over a wide range of launch speeds.

\begin{center}%[ht]
    \begin{minipage}{10cm}  % Adjust the width as needed
        \includegraphics[width=\columnwidth]{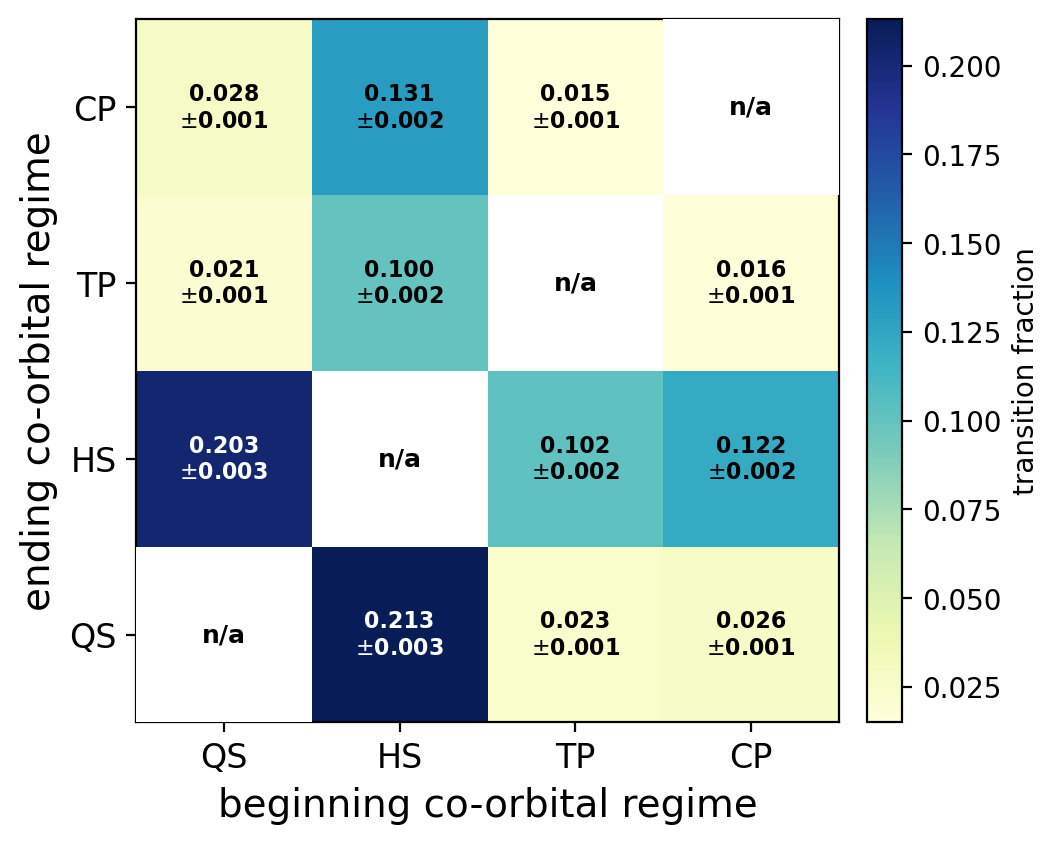}
        \captionof{figure}{
            The fraction of all co-orbital transitions that are from the beginning to ending regime.}
        \label{fig.transition_probabilities}
    \end{minipage}
\end{center}

Transitions between co-orbital regimes, \ie\ where a particle makes a transition without a non-resonant period between the regimes, have been studied for specific transitions and objects \citep[\eg][]{Christou00, Brasser2004, Wajer2010} but we investigated the transition probabilities between all four regimes.  The computed fractions are generally consistent with those already identified in the Sun-Earth system \citep[\eg][]{Brasser2004,Wajer2010}.  We found that they are mostly symmetric with the most statistically significant asymmetry of $\sim2.8\sigma$  between the QS$\rightarrow$HS and HS$\rightarrow$QS transitions (\fig{fig.transition_probabilities}).  Transitions between those two regimes are also the most likely, comprising $\gtrsim20$\% of all jumps between co-orbital states, the most studied and identified, and the one exhibited by \Kamo.  The relatively high probability of HS$\leftrightarrow$CP transitions is expected as they have the same nature as HS$\leftrightarrow$QS transitions, the difference being that CP co-orbitals experience half the resonant angle range of QS co-orbitals. We also expected QS$\leftrightarrow$TP transitions to be difficult because they can occur at non-negligible inclinations.  It would be interesting to analyze these transitions in detail: as a function of inclination, and with respect to eccentricity to determine if they are associated with close approaches to Venus or Mars, and also to the behavior in the argument of pericenter \citep{namouni99}.

%---------------------------------------------
\subsection{Co-orbital steady-state populations}
\label{ss.R+D_co-orbital_steady_state_SFD}

\begin{figure}
    \centering
    \begin{minipage}{15cm} 
    % this figure is made with Earth_co-orbitals.plot_stacked_steady_state_SFD()
        \includegraphics[width=\columnwidth]{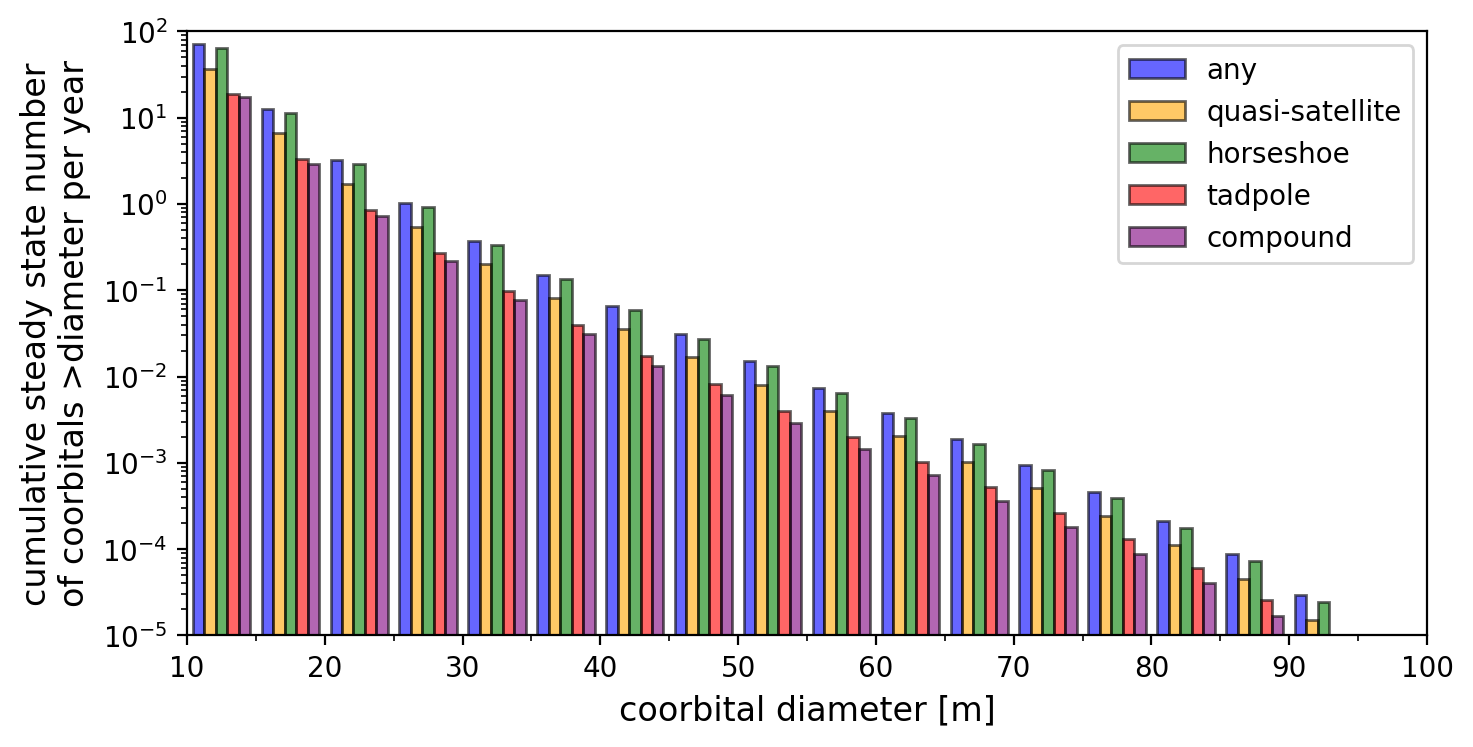}
        \caption{
            The nominal cumulative steady-state SFD of Earth's co-orbitals that originate as lunar ejecta as a function of diameter and their co-orbital regime in bins of $5\meter$ width. 
            }
        \label{fig.lunar_ejecta_coorbital_SFD}
    \end{minipage}
\end{figure}

Our nominal prediction is that there are $\nCumulativeCoorbitalsTenMeterANY$ Earth co-orbitals larger than $10\meter$ diameter in the steady-state population that originated on the Moon's surface (\fig{fig.lunar_ejecta_coorbital_SFD} ~and \tab{tab.Ncoorbital_per_regime}) using the central values for the flux, diameter, and density distribution of objects that impact the lunar surface, the diameter of the crater created by the impact, and the size and speed of the material ejected from the crater (following \citet{Jedicke2025}).  There is an orders-of-magnitude systematic uncertainty on the number, as was found for the population of `temporarily bound objects' with a negative total energy with respect to the geocenter \citep{Jedicke2025}, that is addressed in the following sub-section.  Most of the remaining discussion focuses on the nominal result with the caveat that the systematic uncertainty is large.

The lunar ejecta co-orbital population is primarily composed of objects in the horseshoe regime followed by quasi-satellites, tadpoles and compound objects (\tab{tab.Ncoorbital_per_regime}).  The number of known co-orbital objects in each regime is within a factor of a few of the predicted number of objects with a lunar provenance, and the total number of objects in any co-orbital regime matches the lunar origin prediction to better than 5\%, but this agreement must be spurious because the known population is biased by observational selection effects while the nominal lunar prediction is for the entire population of co-orbitals.

Taken at face value, our nominal lunar provenance model for Earth's co-orbitals predicts that the Moon provides roughly 1/23 ($\sim0.043$) of the steady-state population relative to the main belt.  Our prediction for the number of quasi-satellites $>50\meter$ diameter, the size range corresponding to \Kamo, is $0.008$ (\fig{fig.lunar_ejecta_coorbital_SFD}).  \citet{Fenucci2026} calculated that the number of \Kamo-like quasi-satellites produced specifically by the impact that generated the Giordano Bruno crater would be about $0.042$, a factor of $\sim5$ greater than our value.  While the two calculation methods are not identical we think that they can be compared to one another.  Our value corresponds to the steady-state number due to the formation of craters averaged over a long period of time while their value is for the recent, 1-10$\Myr$ ago formation of the large, and therefore rare, Giordano Bruno crater \citep[\eg][]{Morota2009}.  Our formulation suggests that launching a \Kamo-like object of $50\meter$ diameter requires an $\sim10\km$-scale impactor with an impact frequency of only about 1 per more than 20,000$\Myr$ at the current time.  Thus, our results suggest that \Kamo\ is unlikely to have been ejected by the impactor that formed the Giordano Bruno crater but, if it did so  recently, then we expect that the number of \Kamo-like objects in the current population would be $>0.008$ as predicted by \citet{Fenucci2026}.

Scaling our results for the predicted population of quasi-satellites of main belt provenance from our NEOMOD3 result\footnote{From \tab{tab.Ncoorbital_per_regime} we have 318 quasi-satellites and we use the canonical \citet{Dohnanyi1969} SFD of $\log N\propto H/2$ to scale from the $10\meter$ population ($H=27.75$) to $H=26$ ($\sim35\meter$ diameter\footref{ftn.albedo}).} 
% 318 / 10**((27.75-26)/2) = 42
we predict that there should be $\sim42$ quasi-satellites at the current time with $H<26$ and a main belt provenance, only about 40\% larger than the 30 suggested by \citet{Fenucci2026}.  The difference is likely due to our use of less restrictive ranges in $\aei$ when selecting candidate co-orbitals and our definitions of the co-orbital regimes. Restricting our identified NEOMOD3 co-orbitals to $e<0.2$ and $i<20\arcdeg$ per \citet{Fenucci2026} reduces our predicted co-orbital QS population of main belt provenance to 15.
% the value of 15 provided by EM on 30 Jan 2026 in a comment 

\begin{figure}
    \centering
    \begin{minipage}{15cm} 
        \includegraphics[width=\columnwidth]{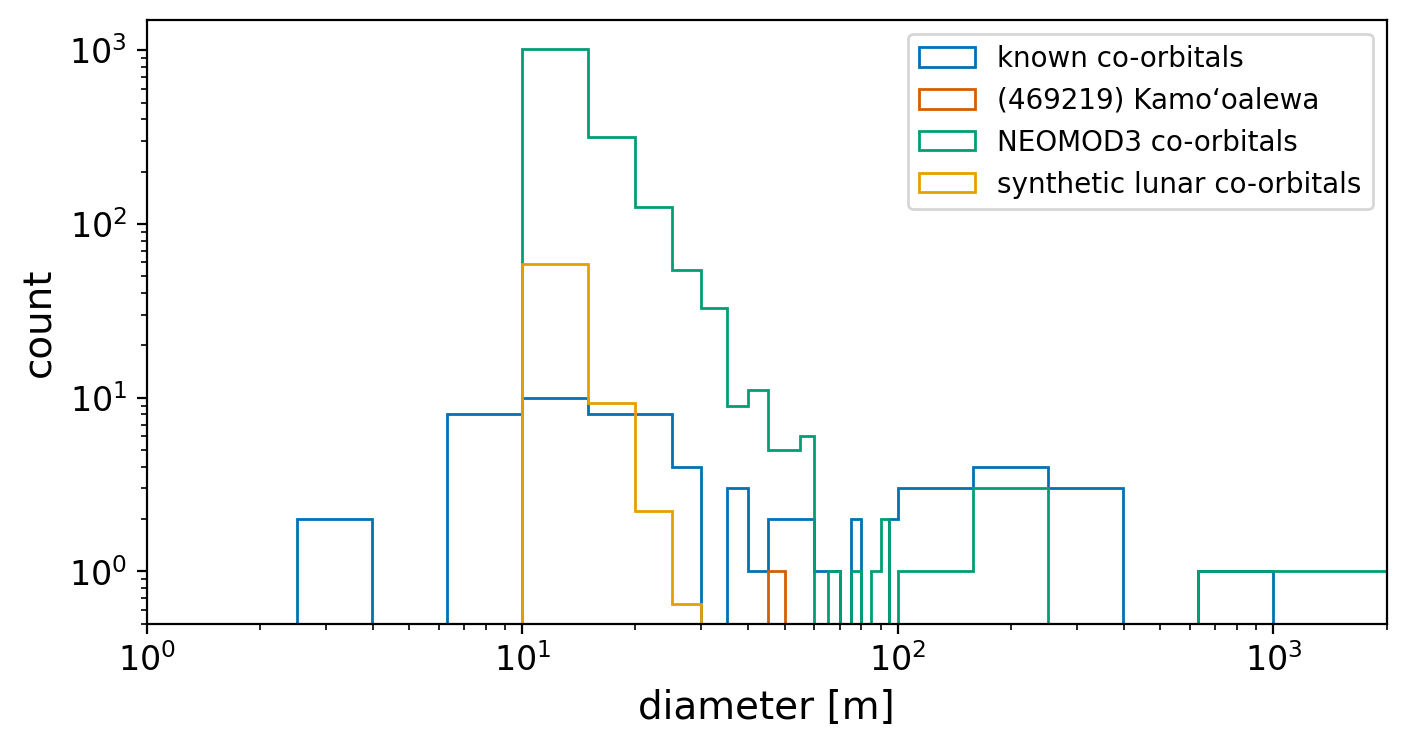}
        \caption{The incremental diameter distributions of the known Earth co-orbitals\protect\footref{ftn.albedo} with \Kamo highlighted, the population of synthetic co-orbitals of main belt provenance suggested by NEOMOD3 \citep{Nesvorny2024-NEOMOD3}, and the nominal distribution of co-orbitals with a lunar origin as calculated in this work.  We have attempted to highlight the diameter range where we have calculated the lunar Earth co-orbital SFD by gerrymandering the bins so there are 5 equal-size logarithmic bins in the range from 1-10$\meter$, 20 bins of $5\meter$ width in the 10-100$\meter$ range, and 6 equal-size logarithmic bins in the range from 100-2000$\meter$.}
        \label{fig.coorbital_diameter_distributions}
    \end{minipage}
\end{figure}

\citet{Fenucci2026} simulated the discovery efficiency of asteroid surveys up to 2022 and found that they would have discovered about 14\% of the quasi-satellites with $H<26$.  The quasi-satellite discovery efficiency would increase slowly with time because they have long synodic periods so we would expect there to be about 6 
% = 0.14 * 43
known quasi-satellites with $H<26$ ($\gtrsim20\meter$ diameter\footref{ftn.albedo}) in excellent agreement with the \nKnownCoorbNOWQSHTwoSix\ we identified in the known population (\tab{tab.Ncoorbital_per_regime}, and the 3 identified by \citet{Fenucci2026}).  Once again, the differences in the predicted numbers are likely a consequence of the range of orbital elements for candidate co-orbitals and the co-orbital regime definitions.  We think that the agreement is good given the differences and both results suggest that a lunar source is not necessary to explain the co-orbital population statistics even though some of their colors and spectra suggest otherwise.

Figure~\ref{fig.coorbital_diameter_distributions} highlights how \Kamo\ is unusually large for a co-orbital of lunar origin given our nominal simulations but not unexpected for co-orbitals with a main belt provenance.  It is no surprise that the synthetic populations of main belt and lunar material have similar size-frequency distributions because both populations are fragments produced in powerful collisions between objects of comparable densities and material strengths.  The known co-orbital population is strongly biased by observational selection effects, such that the population is skewed towards large objects and severely incomplete for objects in the decameter size range.

\begin{figure}
    \centering
    \begin{minipage}{10cm} 
        \includegraphics[width=\columnwidth]{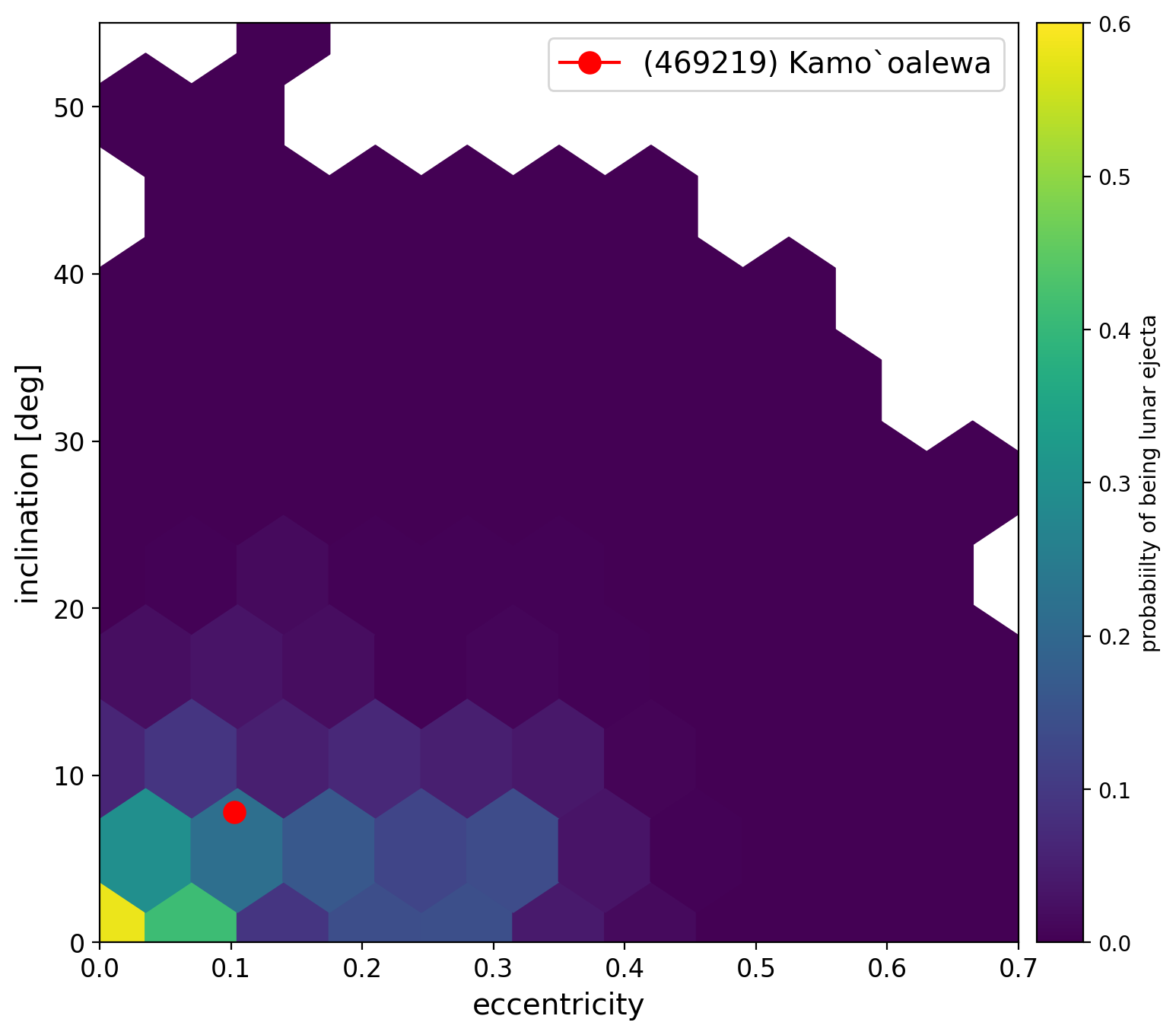}
        \caption{
            The probability that an Earth co-orbital is lunar ejecta versus originating in the main belt or JFC regions assuming our nominal lunar ejecta co-orbital SFD and the NEOMOD3 population (\S\ref{ss.R+D_synthetic_mainbelt_co-orbitals}) as a function of eccentricity and inclination.  White hexes are presumably $\sim0$ but we do not have synthetic objects of either provenance in those bins. The red dot represents \Kamo's orbital elements but note that the probability, like our ejecta simulations, does not account for an object's mass/size.
        }
        \label{fig.lunar_vs_mainbelt_probability}
    \end{minipage}
\end{figure}

\citet{Jedicke2025} suggested that the population of minimoons or, more generally, objects temporarily bound to Earth, could be used to constrain parameters that determine the size-frequency distribution of objects that originated on the Moon.  The SFD of these objects is particularly sensitive to the crater scaling relation, the crater ejecta size distribution, and the relationship between ejecta mass and speed, so that determining whether they are of lunar origin and then measuring their SFD could shed light on all those processes.  In the same way, the SFD of lunar co-orbitals in the regimes studied in this work would be even more useful given that they are larger, more common, and remain visible for longer durations.

\begin{figure}[ht]
    \centering
    \begin{minipage}{10cm} 
        \includegraphics[width=\columnwidth]{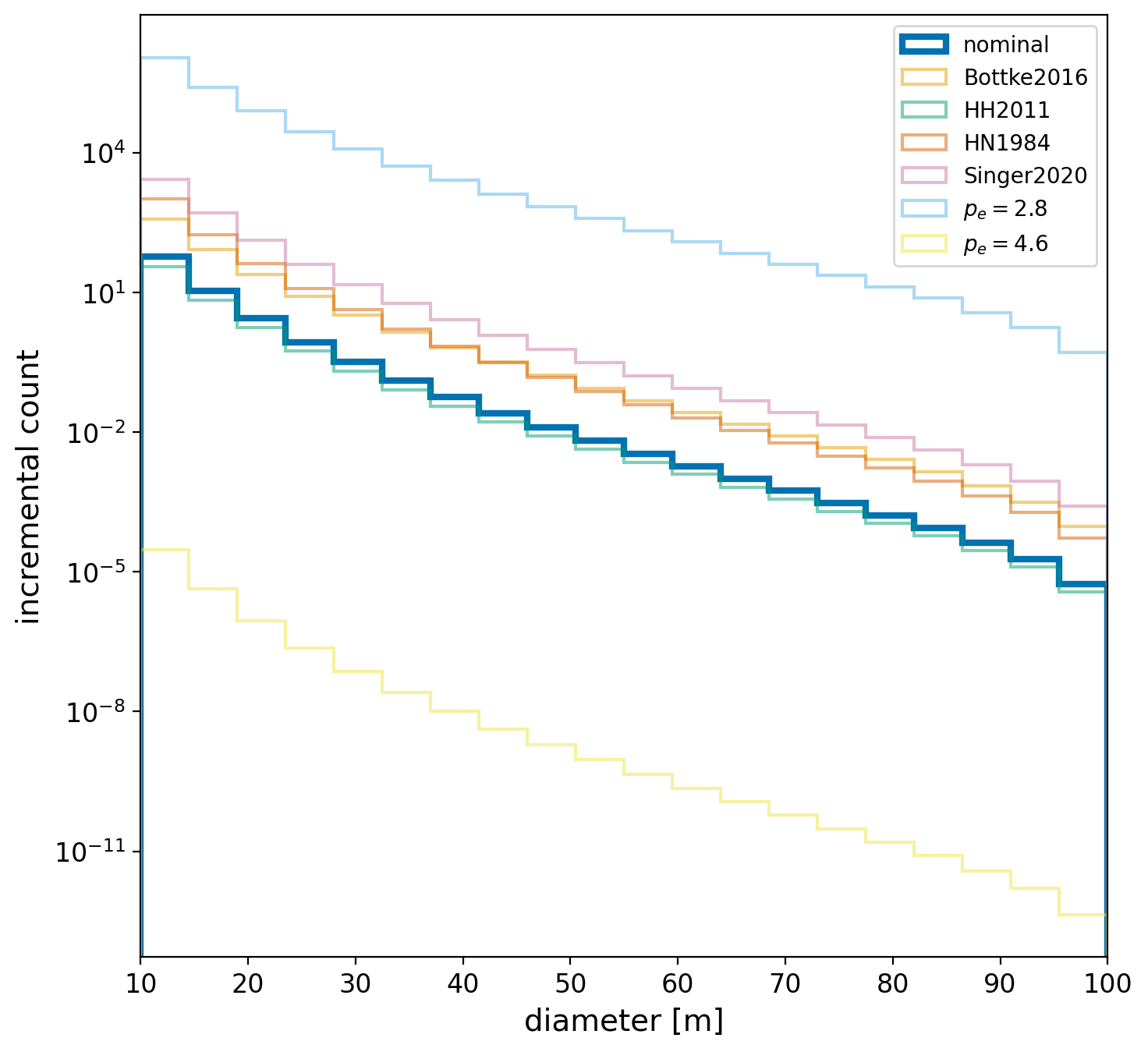}
        \caption{
            Our nominal steady-state SFD of Earth co-orbitals with a lunar provenance and the SFD using four other crater-scaling relations and $\pm1\sigma$ values on the ejecta's SFD where the number of ejecta with diameters larger than $d$ is $N(>d) \propto d^{-p_e}$. The nominal scaling relation is from \citet{Collins2005-craterScaling}, 
            Bottke2016 = \citet{Bottke2016-LPSC-CraterScalingLaws}, 
            HH2011     = \citet{Housen+Holsapple2011-EjectaFromImpactCraters},
            HN1984     = \citet{Horedt1984-SixCraterScalingLaws}
            Singer2020 = \citet{Singer2020-EjectaBlockSizes}.
            The nominal value of $p_e=3.7$ was used in \citet{Jedicke2025} which was an average of values from \citet{Bart2010-BouldersEjectedFromLunarCraters}.
        }
        \label{fig.systematics}
    \end{minipage}
\end{figure}

The probability that a co-orbital is of lunar provenance ($p_{lunar}$, \fig{fig.lunar_vs_mainbelt_probability}) is a function of eccentricity and inclination
\begin{equation}
    p_{lunar}(e,i) = \frac{n_{lunar}(e,i)}{n_{lunar}(e,i) + n_{MB}(e,i)}
\end{equation}
where $n_{lunar}$ and $n_{MB}$ are number of co-orbitals of lunar and main belt origin using our nominal lunar and NEOMOD3 results respectively.  The maximum probability is about \LunarCoorbitalPercentMAX\% for $e\sim0$ and $i\sim0\arcdeg$, decreasing to effectively 0\% for $e\gtrsim0.45$ or $i\gtrsim25\arcdeg$.  These probabilities should be viewed with suspicion because of the tremendous systematic uncertainties in the lunar co-orbital population (\S\ref{ss.SystematicUncertainties}) but the trend will not be affected.  Thus, our results predict that if there are co-orbitals of lunar origin they are significantly more likely to be objects with low eccentricity and inclination.  The nominal probability that \Kamo\ is of lunar origin is \LunarCoorbitalPercentKamo\% based only on its orbital elements, not accounting for its size as discussed above.

The disagreement between our nominal prediction for the population of co-orbitals with a lunar origin and the spectroscopic evidence that some of these objects have surfaces consistent with the lunar surface suggests that there are errors in our assumptions and/or the values we employed in \eg, the crater scaling relations or the ejecta's size-speed relationship.  Alternatively, the association of co-orbital spectra with lunar material may be called into question as even \Kamo\ has been spectroscopically connected with the inner main belt Flora family \citep{Zhang2024}.  Thus, we encourage more characterization of Earth's co-orbitals to clarify the provenance of these objects.

%---------------------------------------------
\subsection{Systematic uncertainties}
\label{ss.SystematicUncertainties}

\citet{Jedicke2025} performed a systematic analysis of most of the important parameters employed in their calculation of the steady-state SFD of temporarily bound objects with a lunar provenance and found that there were many orders of magnitudes systematic uncertainty in their model.  Given that this analysis implements the same methodology but applied to Earth's co-orbitals, there are also many orders of magnitude uncertainty in our nominal results (\fig{fig.systematics}).  Rather than re-perform their study we provide a few examples of the impact of a few of the important input parameters on our results.

The SFD of the lunar ejecta in the cratering event induces $\pm5$ orders of magnitude range in the predicted number of Earth co-orbitals $>10\meter$ diameter of lunar origin extending from $\sim10^{-5}$ to $>10^5$  (\fig{fig.systematics}).  The choice of crater-scaling relation, the functional form of the relationship between an impactor's properties and the final diameter of the crater, changes the predicted number of ejecta by 2 orders of magnitude while our nominal prediction lies on the lower end of co-orbital SFDs produced by five relations.  

The most surprising aspect is that our nominal prediction produces values that are not outrageously out-of-order with the observed population and this suggests that the formal uncertainties on the input parameters may be over-estimated. It also suggests that determining the SFD of co-orbitals with a lunar spectra could assist the effort to constrain our understanding of the crater formation process including crater-scaling relations, the SFD of lunar ejecta, and their size-speed relationship.

%##################################################
\section{Conclusions}

We have calculated the steady-state number of Earth co-orbitals ejected by impacts on the lunar surface by asteroids and comets, performed an independent determination of the number of known co-orbitals, and estimated the number of co-orbitals expected from the main belt utilizing a realization of the population of near-Earth objects with a main belt and cometary provenance.  

The algorithm used to classify co-orbital motion allows for the detection of co-orbital transitions.  The relative fraction of transitions between the four co-orbital regimes matches our expectations with QS$\leftrightarrow$HS transitions being the most common and QS$\leftrightarrow$TP being uncommon.  The computed fractions are consistent with the ones observed so far in the Sun-Earth system.

Our calculation of the expected number of co-orbitals that originated as lunar ejecta has approximately $\pm5$ orders of magnitude systematic uncertainty due to uncertainties on the model's input parameters such as the crater-scaling relation, the size-frequency distribution of the ejecta liberated by the impact, and the size-speed relationship of the ejecta.  Ignoring both the statistical and systematic uncertainties, our nominal prediction is that there are \nCumulativeCoorbitalsTenMeterANY\ Earth co-orbitals larger than $10\meter$ diameter with a lunar provenance at any time. 

This number is spuriously close to the \nKnownCoorbNOWANYTenMeter\ known Earth co-orbitals in the same size range.  The value is `spurious' because the known population in this size range is incomplete due to observational selection effects which make co-orbitals difficult to detect, especially the smaller ones.  Furthermore, the main belt must also supply the Earth co-orbital population and our results suggest that there are about \nNEOMODcoorbAPPROX\ co-orbitals of a main belt origin, more than $20\times$ our nominal result for the lunar origin hypothesis.  Thus, while the Moon can generate Earth's co-orbitals it is not necessary to invoke a lunar provenance to explain the known population.

Our classification of the co-orbital regimes of objects with a main belt provenance suggests that about half of them are in tadpole, Trojan-like, orbits, but only 2 are generally recognized while we identified an additional 3 objects.  This is most likely an observational selection effect because detecting small Earth Trojans is complicated by their large phase angles and small solar elongations.  A population of Earth Trojans would be a valuable source of low $\Delta v$ asteroids for mining or scientific missions and we predict that the Vera Rubin Observatory will discover many of these objects during a twilight survey.

Assuming the co-orbitals have a main belt provenance predicts that they are dominated by objects from the inner main belt and are therefore primarily taxonomically in the S-complex.

The difference between the number of co-orbitals expected from the main belt and our nominal lunar model, and the uncertainly on the latter population, make it all the more surprising that the colors of some co-orbitals seem to be more consistent with the Moon's surface than common asteroid taxonomies.  If this pattern continues for future spectroscopic observations of the co-orbital population it will indicate that impacts are more efficient at launching ejecta than the nominal parameters employed in our analysis suggest.  Measuring the size-frequency distribution of co-orbitals of lunar origin could then constrain impact processes such as crater-scaling relations, ejecta SFDs, and the correlation between ejecta size and speed.

The probability of identifying co-orbitals with spectra similar to the Moon should depend on an object's eccentricity and inclination with main belt objects dominating the high $e$ and $i$ orbital element phase space and lunar ejecta predominant when $e\rightarrow0$ and $i\rightarrow0\arcdeg$.  Disentangling the relative contribution of the co-orbital sources will then require debiasing the observations as a function of their orbital elements.

% does the comparison wrt the other works make sense? maybe the conclusions are too sensitive to the assumptions taken in terms of time step and duration.

% Future: same analysis for the Main Belt, in particular the propagation from NEOMOD3 must be longer and analyzed in the same way (i.e., estimate of durations and transitions).

% Future: link between given co-orbitals and given captures
% IMPORTANT FOR NEXT PAPER :) DON'T REMOVE
%\todo{Elisa Maria}{can I find a link between given co-orbitals and given captures? I would say to try to find only some specific cases. From \fig{fig.coor_capt} it seems that there might be connection between capture and horseshoe, but I should find better and more cases. At the moment I am not able to find them. See also the oscillation in $a$ for the capture ones (figures in 100/capture\_coor).

%##################################################

\section*{Acknowledgments}

We are grateful to Gabriele Merli and Tommaso del Viscio who helped implement the integrations on the Universit\`a di Pavia EOS HPC cluster.
E.M. Alessi was supported by the Fondazione Cariplo and the short-term mobility programme of CNR. 
Uncertainties on all the efficiencies and fractions in histograms and elsewhere were calculated using the technique and software developed by \citet{Paterno2004-EfficiencyUncertainty}.
This work made extensive use of the NumpPy \citep{NumPy}, SciPy \citep{SciPy},  Astropy \citep{Astropy2013,Astropy2018,Astropy2022} and Mathplotlib \citep{Hunter2007} packages.
We thank Victoria Coleman for proofreading the original manuscript and Mikael Granvik and an anonymous reviewer for their careful reviews and constructive suggestions to improve this work.

\bibliographystyle{abbrvnat}
\bibliography{references}

\newpage

%##################################################
\appendix
\setcounter{table}{0}
\renewcommand{\thetable}{A.\arabic{table}}

\section*{Appendix - Known Earth co-orbitals}
\label{s.AppendixTable}

\begin{longtable}{llrll}
\caption{
    Known asteroids that are currently Earth co-orbitals sorted by their designations. JD is the nearest Julian date to the current date that the object had semi-major axis $a=1$. The eccentricity, $e$, and inclination, $i$ are provided at JD. \Kamo\ is 2016~HO$_3$. The bold face initialism in the last column represents the co-orbital regime that the object currently occupies and the other initialisms indicate others that the object experiences during the period of the study from 1600-2500.
    \label{tab.Earth_co-orbitals_known_1600_2500_NOW}
    }\\
\toprule
JD         & $e$  & $i$ [deg] & designation  & co-orbital regime \\
\midrule
\endfirsthead

\multicolumn{5}{c}%
{{\bfseries Table \thetable\ continued from previous page}}\\
\toprule
JD         & $e$  & $i$ [deg] & designation  & co-orbital regimes \\
\midrule
\endhead

\midrule
\multicolumn{5}{r}{{Continued on next page}}\\
\endfoot

\bottomrule
\endlastfoot
2415948.96 & 0.51 & 19.79 & 1986 TO          & {\bf CP} \\
2447165.65 & 0.35 & 33.14 & 1998 UP$_1$      & {\bf CP} \\
2452475.70 & 0.23 &  1.95 & 2000 PH$_5$      & {\bf HS} + CP \\
2452001.06 & 0.17 &  4.67 & 2001 GO$_2$      & {\bf HS} + QS \\
2452550.15 & 0.01 & 10.74 & 2002 AA$_{29}$   & {\bf HS} \\
2451752.57 & 0.04 &  4.29 & 2003 YN$_{107}$  & {\bf HS} + QS \\
2464634.68 & 0.14 & 13.66 & 2004 GU$_9$      & {\bf QS} \\
2445884.10 & 0.63 &  2.66 & 2005 UH$_6$      & {\bf TP} + HS \\
2453979.84 & 0.30 & 33.94 & 2005 QQ$_{87}$   & {\bf TP} \\
2447416.15 & 0.38 &  7.11 & 2006 FV$_{35}$   & {\bf QS} \\
2454205.02 & 0.27 &  1.52 & 2009 HE$_{60}$   & {\bf HS} + QS \\
2455112.23 & 0.09 &  6.84 & 2009 SH$_{2}$    & {\bf CP} \\
2458292.93 & 0.37 & 11.67 & 2010 NY$_{65}$   & {\bf CP} \\
2455182.20 & 0.08 & 14.54 & 2010 SO$_{16}$   & {\bf HS} \\
2456673.65 & 0.19 & 20.89 & 2010 TK$_7$      & {\bf TP} \\
2473934.91 & 0.03 &  0.90 & 2011 UD$_{21}$   & {\bf HS} + CP \\
2456338.41 & 0.09 &  0.82 & 2013 BS$_{45}$   & {\bf HS} \\
2522684.91 & 0.45 & 50.06 & 2013 LX$_{28}$   & {\bf QS} \\
2454673.18 & 0.46 & 10.19 & 2014 OL$_{339}$  & {\bf QS} + HS \\
2457295.21 & 0.11 &  9.20 & 2015 SO$_2$      & {\bf HS} + QS \\
2457370.22 & 0.28 &  1.78 & 2015 YA          & {\bf QS} + HS \\
2457381.44 & 0.41 &  2.50 & 2015 YQ$_1$      & {\bf CP} \\
2457734.05 & 0.18 &  7.69 & 2015 XX$_{169}$  & {\bf CP} \\
2458132.16 & 0.13 &  6.39 & 2016 CO$_{246}$  & {\bf CP} \\
2458888.70 & 0.05 & 27.73 & 2016 CA$_{138}$  & {\bf HS} \\
2462506.41 & 0.10 &  7.81 & 2016 HO$_3$      & {\bf QS} + HS \\
2473372.23 & 0.27 &  0.09 & 2016 JA          & {\bf HS} \\
2457810.31 & 0.24 &  2.99 & 2017 DR$_{109}$  & {\bf CP} \\
2457833.43 & 0.27 &  1.73 & 2017 FZ$_{2}$    & {\bf QS} + HS \\
2458746.73 & 0.15 &  8.73 & 2017 SL$_{16}$   & {\bf CP} \\
2458840.54 & 0.21 & 27.21 & 2017 XQ$_{60}$   & {\bf CP} \\
2457039.55 & 0.15 & 22.10 & 2018 AN$_{2}$    & {\bf CP} \\
2456718.64 & 0.03 &  4.40 & 2018 PN$_{22}$   & {\bf HS} + QS \\
2457896.31 & 0.20 &  9.14 & 2018 KS          & {\bf CP} \\
2457367.46 & 0.30 & 19.68 & 2018 XW$_{2}$    & {\bf CP} \\
2437143.40 & 0.04 &  1.84 & 2019 GF$_{1}$    & {\bf HS} \\
2462423.17 & 0.07 &  6.70 & 2019 GM$_{1}$    & {\bf QS} + HS \\
2458763.88 & 0.27 &  7.21 & 2019 SB$_{6}$    & {\bf CP} \\
2459616.43 & 0.28 &  1.66 & 2019 VL$_{5}$    & {\bf HS} \\
2428282.14 & 0.33 &  4.20 & 2019 XS          & {\bf CP} + TP \\
2458851.98 & 0.19 &  0.47 & 2019 YB$_{4}$    & {\bf HS} \\
2459265.91 & 0.16 & 12.76 & 2020 CX$_{1}$    & {\bf CP} \\
2441019.79 & 0.06 &  5.80 & 2020 PP$_1$      & {\bf HS} + QS \\
2459453.40 & 0.13 &  4.95 & 2020 PN$_1$      & {\bf CP} \\
2449742.78 & 0.39 & 13.85 & 2020 XL$_5$      & {\bf TP} \\
2459231.76 & 0.23 & 12.52 & 2021 BA          & {\bf CP} \\
2459359.91 & 0.02 &  2.99 & 2021 LF$_6$      & {\bf CP} + HS + QS \\
2459547.63 & 0.32 &  6.79 & 2021 XV          & {\bf CP} \\
2459875.28 & 0.26 &  2.12 & 2022 UX$_1$      & {\bf HS} + QS \\
2459880.55 & 0.11 &  6.78 & 2022 UU$_8$      & {\bf CP} + QS + HS \\
2459541.38 & 0.17 &  5.63 & 2022 VR$_1$      & {\bf CP} \\
2459936.23 & 0.20 &  2.36 & 2022 YG          & {\bf QS} + CP \\
2443967.77 & 0.45 &  7.00 & 2022 YF$_4$      & {\bf CP} \\
2460021.36 & 0.18 &  2.75 & 2023 FW$_{13}$   & {\bf QS} \\
2460380.09 & 0.15 &  6.87 & 2023 GC$_2$      & {\bf CP} \\
%2473961.53 & 0.28 &  1.25 & 2023 QD$_2$      & {\bf TP} \\
2460173.72 & 0.15 &  4.74 & 2023 QR$_1$      & {\bf CP} \\
2460194.07 & 0.15 &  3.53 & 2023 SP$_2$      & {\bf CP} \\
2460249.14 & 0.22 &  4.49 & 2023 TG$_{14}$   & {\bf CP} \\
2460312.27 & 0.25 &  5.76 & 2024 AV$_2$      & {\bf CP} + TP + HS \\
2463025.05 & 0.35 &  5.94 & 2024 JR$_{16}$   & {\bf TP} \\
2460713.48 & 0.13 &  4.44 & 2025 CJ          & {\bf CP} \\
2461054.12 & 0.11 & 11.87 & 2025 BL          & {\bf QS} + HS \\
2466179.17 & 0.27 & 11.00 & 2025 FP$_4$      & {\bf CP} \\
2455872.27 & 0.17 &  4.54 & 2025 KR$_4$      & {\bf CP} \\
2461238.24 & 0.11 &  1.99 & 2025 PN$_7$      & {\bf QS} + HS \\
2452137.80 & 0.29 & 16.65 & 2025 RB$_7$      & {\bf CP} \\
2460936.40 & 0.10 &  3.55 & 2025 SC          & {\bf QS} + CP + HS \\
\end{longtable}

\begin{table}
\begin{center}
%\begin{tabular}{ |l|l|r|l|l|} 
\begin{tabular}{llrll} 
\hline
JD         & $e$  & $i$ (deg) & designation    & co-orbital regime \\
\hline
2395963.76 & 0.20 &  2.14 & 2002 VX$_{91}$     & HS + CP \\
2507115.77 & 0.33 &  1.39 & 2008 NP$_3$        & CP \\
2524681.26 & 0.03 &  1.78 & 2010 VQ$_{98}$     & HS + QS \\
2386261.35 & 0.04 &  2.31 & 2013 RZ$_{53}$     & HS \\
2429919.90 & 0.02 &  8.20 & 2014 UR            & CP \\
2484521.14 & 0.32 &  0.91 & 2015 XF$_{261}$    & QS \\
2530606.45 & 0.33 &  0.43 & 2018 UC            & HS \\
2433929.49 & 0.12 &  1.70 & 2019 DJ$_1$        & HS + CP \\
2592075.39 & 0.01 &  5.97 & 2019 HM            & HS \\
2493516.17 & 0.15 &  4.10 & 2019 XH$_2$        & CP \\
2436661.34 & 0.21 &  0.94 & 2020 FN$_3$        & HS \\
2482335.34 & 0.02 &  4.73 & 2020 HO$_5$        & CP \\
2517726.61 & 0.46 &  5.97 & 2021 EY$_1$        & CP \\
2418282.93 & 0.26 &  0.82 & 2021 JE$_1$        & HS \\
2531392.10 & 0.02 &  3.09 & 2022 RW$_3$        & QS + HS \\
2477458.62 & 0.04 &  1.86 & 2023 HM$_4$        & CP \\
2415098.45 & 0.19 &  4.72 & 2023 OU            & CP \\
2528508.64 & 0.05 &  8.00 & 2023 RA$_{10}$     & CP \\
2524679.73 & 0.04 &  2.24 & 2023 RX$_1$        & HS \\
2493474.60 & 0.09 &  2.87 & 2023 SO$_{11}$     & HS + QS \\
2408554.79 & 0.34 & 13.97 & 2024 HB            & CP \\
2583032.67 & 0.02 &  0.92 & 2024 PT$_5$        & HS \\
2492505.82 & 0.06 &  4.65 & 2025 DU$_7$        & QS + HS \\
2538363.18 & 0.10 &  1.01 & 2025 EM            & HS + QS \\
2584941.03 & 0.15 & 15.34 & 2025 FP            & CP \\
2498021.74 & 0.18 &  5.46 & 2025 GL            & CP \\
2421472.37 & 0.02 & 21.49 & 2025 QM$_9$        & CP \\
2415098.78 & 0.09 &  3.33 & 2025 UF$_5$        & HS + QS \\
2388483.65 & 0.26 &  2.17 & 2025 UK$_9$        & HS \\
\hline
\end{tabular}
\end{center}
\caption{
    Known asteroids sorted by their designations that are \emph{not} Earth co-orbitals now but are, or will be, co-orbital at some other time in the range of years from 1600-2500. JD is the nearest Julian date to the current date that the object will have, or had, semi-major axis $a=1$. The eccentricity, $e$, and inclination, $i$ are provided at JD.}
\label{tab.Earth_co-orbitals_known_1600_2500_NOT_NOW}
\end{table} 

\end{document}